\documentclass[a4paper,11pt]{article}

\usepackage{jheppub}
\usepackage{color}
\usepackage{amssymb,graphicx}
\usepackage{tikz-feynman} 
\usepackage{amsmath}
\usepackage{epsfig}
\usepackage{lmodern}
\usepackage[normalem]{ulem}
\usepackage{bm}
\usepackage{subfig}
\usepackage{slashed}
\usepackage{physics}
\usepackage{soul,xcolor}
\usepackage{stackengine}

\usepackage{float}
\allowdisplaybreaks

\title{Supernova bounds on new scalars\\ 
from resonant and soft emission}

\author{Edward Hardy,}
\emailAdd{edward.hardy@physics.ox.ac.uk}
\author{Anton Sokolov,}
\emailAdd{anton.sokolov@physics.ox.ac.uk}
\author{Henry Stubbs}
\emailAdd{henry.stubbs@physics.ox.ac.uk}
\affiliation{Rudolf Peierls Centre for Theoretical Physics, University of Oxford, Parks Road, Oxford OX1 3PU, UK}

\abstract{
We study supernova cooling constraints on new light scalars that mix with the Higgs, couple only to nucleons, or couple only to leptons. We show that in all these cases scalars with masses smaller than the plasma frequency in the supernova core are efficiently produced by resonant mixing with the in-medium longitudinal degree of freedom of the photon. The resulting bounds are free from uncertainties associated to the rate of emission of the scalar in nucleon-nucleon scatterings, which would otherwise badly affect the Higgs-mixed and nucleophilic scenarios. Heavier scalars that mix with the Higgs or couple only to nucleons are mostly produced by nucleon bremsstrahlung, and we obtain a conservative approximation for the corresponding rate using a soft theorem. We also analyse the impact of different supernova profiles, nucleon degeneracy, trapping and scalar decays on the constraints.
}

\begin{document}

\maketitle

\section{Introduction}

New scalars that are uncharged under the Standard Model (SM) gauge group and have sub-GeV masses arise in a variety of beyond Standard Model (BSM) scenarios. They can act as mediators of interactions between dark matter and the SM \cite{Krnjaic:2015mbs,Knapen:2017xzo} or be dark matter themselves \cite{Silveira:1985rk,McDonald:1993ex}, they could play a role in the generation of neutrino masses \cite{Dev:2017dui}, and they might even solve the electroweak hierarchy problem \cite{Graham:2015cka,Flacke:2016szy}. The only renormalisable couplings that such a scalar, which we denote $\phi$, can have to the SM involve a Higgs-portal. The corresponding interaction terms in the Lagrangian are proportional to $H^{\dag} H$, where $H$ is the SM Higgs field. As a result, it is natural to consider theories in which $\phi$ mixes with the Higgs, 
which leads to interactions between $\phi$ and the elementary SM fermions $f$ of the form 
\begin{equation} \label{eq:higgs_portal}
\sin\theta\, \frac{m_f}{v} \phi \bar{f}f~,
\end{equation}
where $m_f$ is the fermion's mass, $v$ is Higgs' vacuum expectation value, and $\theta$ is the mixing angle (the resulting coupling strength of $\phi$ with nucleons $N$ is approximately $8\times 10^{-4}\sin\theta$, which is much smaller than $\sin\theta \,m_N/v$ due to the QCD contribution to nucleons' masses). Meanwhile, more complicated theories can involve scalars that have couplings similar to Eq.~\eqref{eq:higgs_portal} but only to leptons or nucleons (particular realisations can be found in e.g.  Ref.~\cite{Batell:2016ove}). 
For example, ``leptophilic" scalars often appear as mediators in models relevant to direct detection searches for low mass dark matter \cite{Hochberg:2015pha,Hochberg:2015fth}. 
Scalars with interactions of the general form of Eq.~\eqref{eq:higgs_portal} are an important target of the extensive experimental programme searching for new light particles including the planned experiment SHiP \cite{Lanfranchi:2243034}.

Observations of stars lead to strong bounds on the couplings of low-mass BSM particles to the SM \cite{Sato:1975vy,Dicus:1978fp,Vysotsky:1978dc,Dicus:1979ch,Frieman:1987ui,Turner:1987by,Ellis:1987pk,Raffelt:1990yz,Cullen:1999hc}. In particular, agreement between theoretical predictions of stellar evolution and that inferred from observations would be disrupted by the existence of a new light degree of freedom that is sufficiently strongly coupled to the SM since this provides an additional energy loss or transport channel. A given type of star only constrains BSM particles with masses small enough that they can be produced inside the star. Consequently, for new particles with masses greater than approximately $100~{\rm keV}$ (somewhat larger than the temperature in the cores of horizontal branch stars and red-giants) kilonovae and supernovae are the only relevant systems. Specifically, core-collapse supernovae are among the most extreme known physical processes with typical temperatures and densities in their cores exceeding $10~\text{MeV}$ and $10^{13}~\text{g/cm}^3$ respectively. These conditions are ideal for the copious production of BSM particles with masses up to approximately $100~\text{MeV}$. Even for small couplings to the SM, such particles could dramatically affect the dynamics of a supernova, especially because energy loss from the supernova by SM processes is inefficient.

In this paper we reanalyse the supernova constraints on CP-even scalars, with the aim of obtaining limits that are as robust as possible and with a detailed understanding of the remaining uncertainties. 
This is worthwhile because observations of SN~1987A exclude a sizable region of the parameter space of such scalars that is both otherwise unconstrained and relevant for well-motivated theories, e.g. in which the scalar is a mediator to dark matter \cite{Knapen:2017xzo}. 
We primarily present our analysis in terms of the Higgs-mixed pattern of couplings of Eq.~\eqref{eq:higgs_portal}. However, after straightforward adaptations, we also obtain limits on scalars that couple only to nucleons. Additionally, we give results for low mass scalars, $m_\phi\lesssim 10~{\rm MeV}$, that couple only to leptons, leaving the case of heavier leptophilic scalars for future work.

For scalars with relatively small mass, $m_\phi\lesssim 10~{\rm MeV}$, we obtain limits by considering production by resonant mixing with the in-medium longitudinal photon. Provided the plasma frequency $\omega_p > m_{\phi}$ there is always a frequency at which such a resonance occurs. As we will see, the resulting rate at which $\phi$ particles are produced is independent of the sizable uncertainties on the $NN\rightarrow NN\phi$ cross section that would otherwise plague the Higgs-mixed and nucleophilic cases.

Meanwhile, for $m_\phi > \omega_p$ the mixing of the scalar and the photon is negligible and, for the Higgs-mixed and nucleophilic cases, it is expected that $\phi$ is dominantly produced directly by nucleon bremsstrahlung. 
The rate at which this occurs can be estimated in various ways. 
For the similar case of axions, the one-pion-exchange (OPE) approximation \cite{Friman:1979ecl} has often been used \cite{Iwamoto:1984ir,Brinkmann:1988vi}, however this misses potentially important effects, e.g. due to two pion exchange. 
Instead of the OPE approximation, Ref.~\cite{Hanhart:2000ae} made use of soft theorems to relate the axion emission rate to experimental nucleon-nucleon scattering data and this resulted in an order of magnitude reduction in the emissivity. 
Similarly, in Ref.~\cite{Chang:2018rso} the OPE rate was corrected by various phenomenological factors to account for higher order scattering effects, and Ref.~\cite{Carenza:2019pxu} studied corrections to the OPE approximation in detail leading to results comparable to those obtained from the soft approximation. Analogous studies have been carried out for the case of dark photon production in supernovae, see e.g. Ref.~\cite{Rrapaj:2015wgs}. Given this, we choose to calculate $\phi$'s production rate using the soft approximation since we expect this to lead to conservative results.  We also take into account the effect of nucleon degeneracy on bremsstrahlung production in the supernova. Despite {\it a priori} being potentially important because the nucleons are partially degenerate in the core, this turns out to only have a minor effect on our final constraints.

Previous work studying supernova bounds on scalars includes the early work of Ref.~\cite{Ishizuka:1989ts} (related analysis of a saxion has also been carried out \cite{Arndt:2002yg}). More recently, Refs.\cite{Diener:2013xpa,Krnjaic:2015mbs} obtained constraints using the OPE approximation. These bounds were reanalysed in \cite{Dev:2020eam}, which pointed out an important cancellation in the OPE approximation that means that emission of the scalar from the internal pion propagator dominates for scalars with small masses. 
Subsequently \cite{Balaji:2022noj} again used the OPE but studied the impact of decay and reabsorption in detail and analysed the impact of different supernova profiles. Meanwhile, resonant production of new light particles was first proposed in the context of dark photons \cite{Redondo:2008ec}, revised in Refs.~\cite{Redondo:2013lna,An:2013yfc}, and applied to supernovae in \cite{Chang:2016ntp,Hardy:2017}. Resonant production of scalars in horizontal branch stars and red-giants was analysed in \cite{Hardy:2017}, see also \cite{Bottaro:2023gep,Yamamoto:2023zlu}. Similar plasma mixing effects have also been considered in other contexts involving new light particles, e.g.  \cite{Raffelt:1987im,Dvorkin:2019zdi,Gelmini:2020xir,DeRocco:2022rze,Brahma:2024vxb}. Recently, Ref.~\cite{Springmann:2024mjp} studied finite temperature and density corrections to axion production in supernovae, and corrections to the OPE approximation, using the real-time formalism which we also employ.

The remainder of our work is structured as follows. In Section~\ref{sec:supernovae} we summarise the current understanding of the relevant supernova physics, how observations of SN~1987A lead to constraints on BSM particles, and the reference supernova model that we use; although this material is mostly standard it is crucial for accurately interpreting our final constraints. In Section~\ref{sec:production} we calculate the production rate of scalars in the supernova environment from resonant and soft emission. In Section~\ref{sec: Trapping and decay} we analyse the effects of reabsorption and decays of the scalars within a supernova. We present our results for Higgs-mixed, nucleophilic, and leptophilic scalars in Section~\ref{sec:results},
and we discuss our work and possible extensions in Section~\ref{sec:discussion}. Further details are provided in Appendices.

\section{Supernovae}\label{sec:supernovae}
\subsection{Raffelt criterion}\label{rfc}

Identifying the observational implications of BSM particles for the evolution of a supernova core and the aftermath of the explosion is not straightforward. 
Almost all existing observational data on supernovae comes from the electromagnetic radiation of the supernovae ejecta and remnants. The properties of this radiation are related to the evolution of the supernova core, and therefore also to the effects of BSM particles produced in the core, only in an indirect way subject to many modelling uncertainties. 
One notable exception is SN~1987A -- a type II supernova in the Large Magellanic Cloud -- the neutrino signal from which was detected by multiple observatories~\cite{Bionta:1987qt, Kamiokande-II:1987idp, Alekseev:1987ej}. Contrary to the electromagnetic signal, the neutrino signal allows the supernova core to be studied directly, because neutrinos that are emitted from the core travel through the outer layers of the star unimpeded. 

An important feature of the neutrino signal from SN~1987A is its duration of roughly $10\,\text{s}$, which is in agreement with the most established theory for the core-collapse supernova explosion, involving the so-called delayed neutrino mechanism~\cite{Bethe:1985sox}. In this, the stalled supernova shock is revived by neutrino energy deposition within a second after the core bounce, thus drastically diminishing the subsequent infall of matter onto the core and potentially preventing it from collapsing into a black hole. Instead, there can remain a new-born protoneutron star (PNS) that then gradually cools down by emitting neutrinos with energies of order $10~\text{MeV}$ from its surface over a timescale of order $10\,\text{s}$. It has been argued that the spectrum and duration of the observed neutrino burst from SN~1987A is consistent with a neutrino-driven explosion~\cite{1987ApJ...318L..63B, 1988ApJ...334..891B}.

If the observed neutrino burst is indeed explained by the cooling of the PNS, this signal can be used to put constraints on hypothetical very weakly interacting BSM particles~\cite{Raffelt:1987yt, Turner:1987by,Raffelt:1990yz}. Such particles can carry away energy similarly to neutrinos, contributing to an additional cooling of the star and therefore shortening the neutrino burst. A simple and intuitive criterion has been proposed: new particles cannot carry away more energy than neutrinos do. The total energy emitted in neutrinos can be estimated from both the observed neutrino signal and simulations to be approximately $\, (2~\text{--}~3)\times 10^{53}~\text{erg}$. This leads to an upper bound on the average luminosity of the protoneutron star in BSM particles of
\begin{eqnarray}\label{raff}
	L_{\text{BSM}}\, \lesssim \, \, (2~\text{--}~3)\times 10^{52}~\text{erg}/\text{s} \, ,
\end{eqnarray}
which is known as the Raffelt bound (for definiteness, our subsequent results are obtained for $3\times 10^{52}~\text{erg}/\text{s}$; we find that the uncertainty from Eq.~\eqref{raff} is small compared to other sources of uncertainties). 

The Raffelt bound leads to strong constraints on the couplings of BSM particles to the SM. These are particularly stringent because sufficiently weakly coupled BSM particles can free-stream directly out of the whole interior of the PNS whereas, famously, neutrinos are  trapped and thus emitted solely from the surface. Even BSM particles with tiny couplings to the SM would therefore drain energy much more efficiently than neutrinos. 
Increasing the coupling of a BSM particle to the SM leads, at some point, to a decrease in its luminosity $L_{\text{BSM}}$ due to trapping as bulk emission gradually gives way to surface emission~\cite{Burrows:1990pk}.\footnote{In specific cases~\cite{Caputo:2021rux, Caputo:2022rca}, it was found that the emission of BSM particles in the trapping regime does not exactly correspond to the surface emission, but is rather dominated by an extended shell that gets farther from the centre of the PNS when the coupling is increased; moreover, it was found that the overall emission does not always follow the Stefan-Boltzmann law~\cite{Caputo:2021rux, Lai:2024mse}.} As a result,  Eq.~\eqref{raff} is satisfied again for sufficiently strong couplings and the constraints from SN~1987A cover only a band in the coupling-mass parameter space of a given candidate BSM particle. 
Couplings corresponding to the lower region of such a band (and smaller) lead to the free-streaming regime while those corresponding to the upper part (and larger) lead to the trapping regime.

Due to the complexity of the physics involved, a reliable modelling of the supernova explosion, including the PNS cooling phase, requires numerical simulations. Such a numerical approach is therefore also needed to verify the Raffelt criterion. Early support for this criterion was provided by the spherically symmetric simulations of Refs.~\cite{Burrows:1988ah, Mayle:1987as, Mayle:1989yx}. However, these were performed assuming that the stalled supernova shock is indeed successfully revived by neutrinos, which required the energy deposition by neutrinos into the stalled shock to be artificially increased (this modification seems essential to model spherically symmetric neutrino-driven supernovae). 
As a result, such simulations are not self-consistent, which casts doubt on the strength of evidence they provide for the Raffelt criterion~\cite{Bar:2019ifz}.  
Additionally, a recent extensive study~\cite{Fiorillo:2023frv} has questioned whether the observed neutrino signal is consistent with the predictions of the standard neutrino-driven core-collapse supernova models. In particular, this reference argued that the late-time events observed in K-II and BUST (Baksan) detectors are difficult to reconcile with modern spherically symmetric simulations. 
A possible solution to the need for artificial shock revival has been found in 3-dimensional simulations: it turns out that relaxing the constraint of spherical symmetry allows this to occur naturally~\cite{Lentz:2015nxa, Muller:2017hht, Burrows:2019zce, Vartanyan:2021dmy}. It has been hypothesized that going to 3 dimensions can also help to account for the observed late-time neutrino events, which might be associated with late-time asymmetric accretion onto the PNS~\cite{Fiorillo:2023frv}. However, as yet no such self-consistent 3-dimensional simulations have been performed including additional energy loss due to BSM particles.

There are additional uncertainties in the trapping regime. In this case, the reabsorption of BSM particles contributes to energy transfer within the PNS, and the effect of this on the evolution of the PNS and the neutrino signal is hard to predict analytically and even numerically. In particular, essentially 3-dimensional processes such as PNS convection and accretion onto the PNS that play an important role for neutrino emission are likely to be affected by the additional energy transport. Moreover, the reabsorption of BSM particles could influence the earlier stages of the supernova such as the core-collapse phase, potentially modifying the input for simulations of the PNS evolution. 
Early spherically symmetric numerical simulations~\cite{Burrows:1990pk} suggest that in the trapping regime the neutrino signal is indeed shortened if the BSM particle luminosity from the PNS is sufficiently large, but a detailed 3-dimensional study is currently lacking.

It is also important to note that there has been a recent reanalysis of the Raffelt criterion~\cite{Bar:2019ifz}, which argued that SN~1987A could have been ignited by the so-called collapse-induced thermonuclear explosion (CITE) mechanism~\cite{Burbidge:1957vc, Fowler:1964zz, Hoyle:1960zz,Kushnir:2014oca, Blum:2016afe, Kushnir:2015mca, Kushnir:2015vka} instead of the delayed neutrino mechanism. CITE would lead to the formation of a black hole, and the later ($t\gtrsim 4\,$s) part of the neutrino signal could originate from an accretion disk formed shortly after this. The neutrino luminosity of such an accretion disk would not be strongly affected by BSM particle emission, dramatically weakening the resulting constraints~\cite{Bar:2019ifz}. 
Interestingly, the CITE scenario could explain both the above-mentioned ``anomalous" late-time neutrino events as well as the gap observed in the neutrino signal after the initial $t \simeq 2\,$s burst. On the other hand, the existing (limited) literature suggests that CITE ignition requires a very special choice of supernova progenitor, with tuned initial density and composition profiles that do not arise naturally in self-consistent stellar evolution codes~\cite{Gofman:2018ldp}.  Still, it is known that the progenitor star of SN~1987A had unusual features that have resisted robust explanation within  stellar evolution models for decades~\cite{1992PASP..104..717P, Utrobin:2018mjr, Li:2023ulf}; as such, it is not clear whether the main argument against CITE is applicable in this particular case. 
Fortunately, the debate might be resolved by upcoming observations of the SN~1987A remnant, since an observation of the neutron star would rule CITE out. Although as yet there is no evidence for a neutron star, it has recently been argued that the observed features of the remnant are consistent with the neutron star hypothesis~\cite{Page:2020gsx, doi:10.1126/science.adj5796} raising hope of a decisive future observation.

In our present study we adopt the working hypothesis that the delayed neutrino mechanism is the correct description of SN~1987A and impose the Raffelt criterion. We stress that if the tuned progenitors required for CITE are found to arise in the case of the SN~1987A progenitor or a black hole is found within the remnant our results would be greatly affected. 
We also neglect the uncertainties associated to BSM energy transport in the trapping regime and simply assume that any energy drained from the PNS due to BSM particle emission shortens the neutrino signal. Specifically, we base our constraints on the amount of energy emitted in BSM particles from the main neutrino-producing region given by the radii $r\lesssim R_{\nu}$, where $R_{\nu}$ is the neutrinosphere radius.

\subsection{Reference supernova model}

To calculate the rate of BSM particle production and apply the Raffelt bound of Eq.~\eqref{raff}, we need a reference model for SN~1987A. Given the above discussion, this has to be based on the delayed neutrino mechanism and ideally would be 3-dimensional and lead to an explosion. However, despite their sophistication and successes, existing 3-dimensional models are still in their infancy in regard to including all possibly relevant physical effects; no simulation fully takes into account potentially important factors such as magnetic fields, rotation, shell mixing, convective core overshooting~\cite{Temaj:2023nuo}, and progenitor binarity. Moreover, it is hard to compare these simulations to observations due to current large uncertainties especially on the progenitor properties~\cite{Burrows:2020qrp, Burrows:2024pur}.  Given this, for our purposes it is reasonable to consider a simplified, spherically symmetric, supernova model. 
In particular, 
we use the model of Ref.~\cite{Serpico:2011ir}, which was  computed with the spherically-symmetric \textsc{Prometheus}-\textsc{Vertex} simulations~\cite{Rampp:2002bq, Buras:2005rp}. This implements the Lattimer-Swesty equation of state~\cite{Lattimer:1991nc} with a symmetry energy of $29.3$~MeV for nuclear matter, and we assume the value for the bulk incompressibility modulus is $180$~MeV. The resulting profiles of the relevant physical quantities and the neutrinosphere radii are available from the Garching Core-Collapse Supernova Archive~\cite{GarchingArchive} (see also Appendix~\ref{appendix: sn profiles} for details of the profiles). 

The progenitor for SN~1987A is known to have been a star with a mass around $20~M_{\odot}$~\cite{1990ApJ...360..242S, Blinnikov:1999eb, 2011A&A...532A.100U}, however, as mentioned, its exact nature is not fully understood~\cite{1992PASP..104..717P, Utrobin:2018mjr,  Li:2023ulf}. As a rough estimate of the uncertainty associated to the progenitor, as well as to the profiles themselves, we consider four sets of profiles corresponding to progenitors with the masses of $15~M_{\odot}$, $17.8~M_{\odot}$, $20~M_{\odot}$ and $25~M_{\odot}$ (with PNS masses at shock revival of $1.5~M_{\odot}$, $1.6~M_{\odot}$, $1.5~M_{\odot}$ and $1.9~M_{\odot}$, respectively), where the first progenitor is from Ref.~\cite{Woosley:1995ip} and the other three are from Ref.~\cite{Woosley:2002zz}. 
We extract the profiles and the neutrinosphere radii at the moment of the supposed shock revival, i.e. at the very end of the accretion phase when the PNS cooling phase begins. In this way, our results are not affected by the artificial shock revival mechanism or the fact that spherically symmetric profiles are likely to be unreliable after the shock revival (since the subsequent evolution is expected to be fundamentally 3-dimensional~\cite{Burrows:2019zce}). The error introduced by applying the Raffelt bound to the luminosity at the beginning of the cooling phase rather than to the average luminosity during the first seconds of cooling is expected to be minor, at most of order of the uncertainty already present in Eq.~\eqref{raff} and smaller than that from the choice of a progenitor (this is indeed the case if the symmetric profiles are assumed to be accurate after revival).

\section{Production}\label{sec:production}

It can be shown \cite{Laine:2016} that the production rate, $\Gamma_{\text{prod}}(\boldsymbol{k})$, of a weakly interacting real scalar particle $\phi$ with a 4-momentum $K=(\omega, \boldsymbol{k})$ satisfying $\omega^2 = \boldsymbol{k}^2+m_\phi^2$ from a medium in thermal equilibrium is given by 
\begin{equation} \label{eq:GammaProd}
        \Gamma_{\text{prod}}(\boldsymbol{k}) \equiv \dot{\mathcal{F}}_\phi(\boldsymbol{k}) = \frac{1}{2 \omega} \pi^{<}(\omega,\boldsymbol{k}) +\mathcal{O}(g_{f}^4)~,
\end{equation}
where $\mathcal{F}_{\phi}(\boldsymbol{k})$ is the phase space density of the scalars, and $g_{f}$ is the coupling between the scalar and the particles $f$ in the medium. The Wightman function $\pi^<(\omega,\boldsymbol{k})$ is defined by
\begin{equation} \label{eq:pi_J}  
    \pi^{<}(\omega,\boldsymbol{k}) = \int d^4 x \ e^{i K\cdot x}\langle \hat{J}(0)\hat{J}(x)\rangle_\beta~,
\end{equation}
where the subscript $\beta$ denotes a thermal average (i.e. expectation value with respect to the SM bath density matrix) and $\hat{J}$ is the current that couples to the scalar field. In the theories we consider
\begin{equation}\label{smcurrent}
    \hat{J} = \sum_f g_f 
    \bar{\hat{f}} \hat{f}\, ,
\end{equation}
with the sum running over all fermions $f$ that couple to the scalar, each with a strength $g_f$.\footnote{Note that due to PT-invariance of the interactions the Wightman function, Eq.~\eqref{eq:pi_J}, is real.}
An outline of the derivation of Eq.~\eqref{eq:GammaProd} is provided in Appendix~\ref{appendix: Theoretical setup}. To evaluate the two-point function $\pi^{<}$ we turn to the real-time formalism of thermal field theory, an introduction to which can be found in Refs.~\cite{Le_Bellac:1996, Lundberg:2020mwu, Ghiglieri:2020dpq}.

As in the imaginary-time formalism, in the real-time formalism the thermal distribution $e^{-\beta \hat{H}}$ is formally interpreted as a time evolution operator. However, unlike the imaginary-time formalism, in which the path that time takes through the complex plane is parallel to the imaginary axis, the real-time formalism involves a complex time contour that includes the entire real axis. As a result, the real-time formalism allows correlation functions of operators inserted at different real times to be calculated without the need for analytic continuation.  The requirement that the end points of the time contour, $t_i$ and $t_f$, satisfy $t_f = t_i - i\beta$ leads us to the Schwinger-Keldysh time contour, where time proceeds forwards from the past into the future before dropping below the real axis by $\text{Im}\, \Delta t = -\sigma$ to double back on itself, and finally moving down a distance $\beta - \sigma$ into imaginary-time (we take the convention $\sigma=0$). This leads to the concept of contour-ordering, denoted $\langle\ldots\rangle_c$, which generalises the usual time-ordering. It can be shown that the parts of the time contour parallel to the imaginary axis factor out of the partition function, and so only the parts parallel to the real axis need to be kept track of when calculating correlation functions. Therefore,   operators are categorized based on whether they are inserted on the time contour above the real axis, labelled by an index 1, or on the time contour below the real axis, labelled by an index 2. This leads to a propagator that is a $2\times 2$ matrix
\begin{equation}
    D = 
    \begin{pmatrix}
    \langle\phi_1\phi_1\rangle_{c}&\langle\phi_1\phi_2\rangle_{c}\\
    \langle\phi_2\phi_1\rangle_{c}&\langle\phi_2\phi_2\rangle_{c}
    \end{pmatrix}.
\end{equation}
Vertex factors also carry labels to specify whether they are interactions between type-1 fields or type-2 fields (a 3-point vertex now has $2^3=8$ possible index structures).

We can then simply relate the relevant Wightman function to a component of the scalar field's self-energy in the real-time formalism of thermal field theory as $\pi^<(\omega,\boldsymbol{k}) = - i\pi^{12}(\omega,\boldsymbol{k})$ (the factor of $-i$ appears as usual in a relation between a self-energy and a current correlator). This is because fields on the second contour always come after fields on the first contour in the contour ordering prescription, meaning that the 12-component of the self-energy leads to the required ordering of operators in Eq.~\eqref{eq:pi_J} \cite{Ghiglieri:2020dpq}. Actually calculating the self-energy is complicated by the fact that the real-time propagator is a $2\times2$ matrix. This motivates the question of whether we can transform our field basis 
\begin{equation}
\begin{pmatrix}
    \phi_a \\ \phi_b
\end{pmatrix} = U
    \begin{pmatrix}
        \phi_1 \\ \phi_2 
    \end{pmatrix}~,
\end{equation}
such that the propagator matrix $D^{ab}$ simplifies. It is possible to make the propagator diagonal \cite{van_Eijck:1994}, however we choose an alternative field basis known as the $R/A$ basis where the propagator becomes off-diagonal, i.e. $D^{AA} = D^{RR} = 0$. This basis still enjoys many of the advantages of a diagonal basis while also having particularly simple vertex factors and cutting rules \cite{Gelis:1997}. Due to the existence of a spectral representation of the self-energies \cite{Luttinger:1961zz}, the real and imaginary parts of the various components of the self-energies in different bases can be related to one another. For example, we have $i\pi^{12}(\omega,\boldsymbol{k}) = 2n_B(\omega) \text{Im}[\pi^{RA}(\omega,\boldsymbol{k})]$  (formally only valid for $\omega>0$) \cite{Gelis:1997}, which allows us to rewrite the production rate in terms of a quantity in the $R/A$ basis as
\begin{equation}
    \Gamma_{\text{prod}}(\omega)  = -\frac{n_B(\omega)}{\omega} \text{Im}[\pi^{RA}(\omega)]~,
\end{equation}
where $n_B(\omega)$ is the usual Bose occupation factor. Here and in what follows we drop the dependence on $\boldsymbol{k}$ from the production rate and the self-energy, since the medium is isotropic on the scales associated with the inter-particle interactions responsible for BSM particle production and so there is no dependence on the direction of the 3-momentum.\footnote{Our supernova model does not include magnetic fields, which would lead to anisotropies that might affect the thermal field theory calculation~\cite{Brahma:2024vxb}.}

We would now like to understand the different physical contributions to the self-energy $\pi^{RA}(\omega)$ and therefore $\Gamma_{\text{prod}}(\omega)$. There is of course the one-particle-irreducible (1PI) contribution. 
However, there are also contributions that are not 1PI, due to the scalar having a medium-induced mixing with a collective excitation of the plasma known as the ``longitudinal photon" or ``plasmon". The background medium explicitly breaks Lorentz invariance by picking out a preferred frame where the medium is at rest. This allows the photon to acquire a longitudinal polarization state, which has the same quantum numbers as the scalar permitting the two to mix.\footnote{The exception to this is charge parity, which is +1 for the scalar and -1 for the longitudinal photon. However, the medium also explicitly breaks charge conjugation symmetry by having unequal numbers of particles and antiparticles.} These two contributions can be represented diagrammatically as 
\begin{equation}\label{eqn: diagram contributions to self energy}
-i\pi^{RA} = \;
\begin{tikzpicture}[baseline={(0,-0.1)}]
  \begin{feynman}
    \coordinate (i1) at (-2, 0);
    \coordinate (o1) at (2, 0);
    \coordinate (b1) at (0, 0);
    
    \diagram* {
      (i1) -- [scalar] (b1),
      (b1) -- [scalar] (o1),
    };
    
    \filldraw[fill=gray!30] (0, 0) circle (0.7);
    
    \node at (b1) {$\Pi_{\phi\phi}^{RA}$};
  \end{feynman}
\end{tikzpicture}
\;
+\;
\begin{tikzpicture}[baseline={(0,-0.1)}]
  \begin{feynman}
    \coordinate (i2) at (-2, 0);
    \coordinate (b1) at (0, 0);
    \coordinate (b2) at (2.5, 0); 
    \coordinate (o2) at (4.5, 0);
    
    \diagram* {
      (i2) -- [scalar] (b1),
      (b2) -- [scalar] (o2),
    };

    \draw[photon] (b1) -- (b2);
    \draw[photon] (b1) ++(0,-0.07) -- ++(2.5,0);
    \node at (0.9,0.3) {\small A};
    \node at (1.6,0.3) {\small R};
    
    \filldraw[fill=gray!30] (0, 0) circle (0.7);
    \filldraw[fill=gray!30] (2.5, 0) circle (0.7);
    
    \node at (b1) {$\Pi_{\phi \gamma}^{RA}$};
    \node at (b2) {$\Pi_{\gamma \phi}^{RA}$};

  \end{feynman}
\end{tikzpicture}
\;
+\;\mathcal{O}(g_{f}^4)~,
\end{equation}
or in terms of 1PI self-energies as
\begin{equation}
    -i\pi^{RA} = -i\Pi_{\phi\phi}^{RA}+(-i \Pi_{\phi \gamma,\mu}^{RA})D^{AR,\mu\nu}_{\gamma}(-i \Pi_{\gamma\phi,\nu}^{RA})+\mathcal{O}(g_{f}^4)~,
\end{equation}
with $D_\gamma$, corresponding to the double wavy line in Eq.~\eqref{eqn: diagram contributions to self energy}, being the full resummed thermal photon propagator; note that although we draw external legs for self-energy diagrams they are as usual assumed to be amputated. We refer to $\Pi_{\phi\phi}^{RA}$ as the 1PI self-energy, and to $\Pi_{\phi\gamma}^{RA}$ as the 1PI mixing self-energy. In the $R/A$ basis there is only a single non-vanishing vertex assignment in the above diagram, whereas the original $1/2$ basis would have led to a more complicated expression.

In passing, we note that in the case of a dark photon $A$, the fact that the dark photon's interactions are the same as those of the SM photon apart from a universal rescaling by the kinetic mixing $\epsilon$ leads to relations between the mixing self-energy and the photon's self-energy, $\Pi_{AA}=\epsilon \Pi_{A\gamma}=\epsilon^2 \Pi_{\gamma\gamma}$. This results in interference and cancellations between the two terms of Eq.~\eqref{eqn: diagram contributions to self energy} \cite{Redondo:2013lna,An:2013yfc,Chang:2016ntp,Hardy:2017} referred to as screening. However, in our case the self-energies are unrelated and hence we are able to split Eq.~\eqref{eqn: diagram contributions to self energy} into two physically distinct production channels, which we can consider independently from each other.

\begin{figure}
    \centering
    \includegraphics[width=0.7\linewidth]{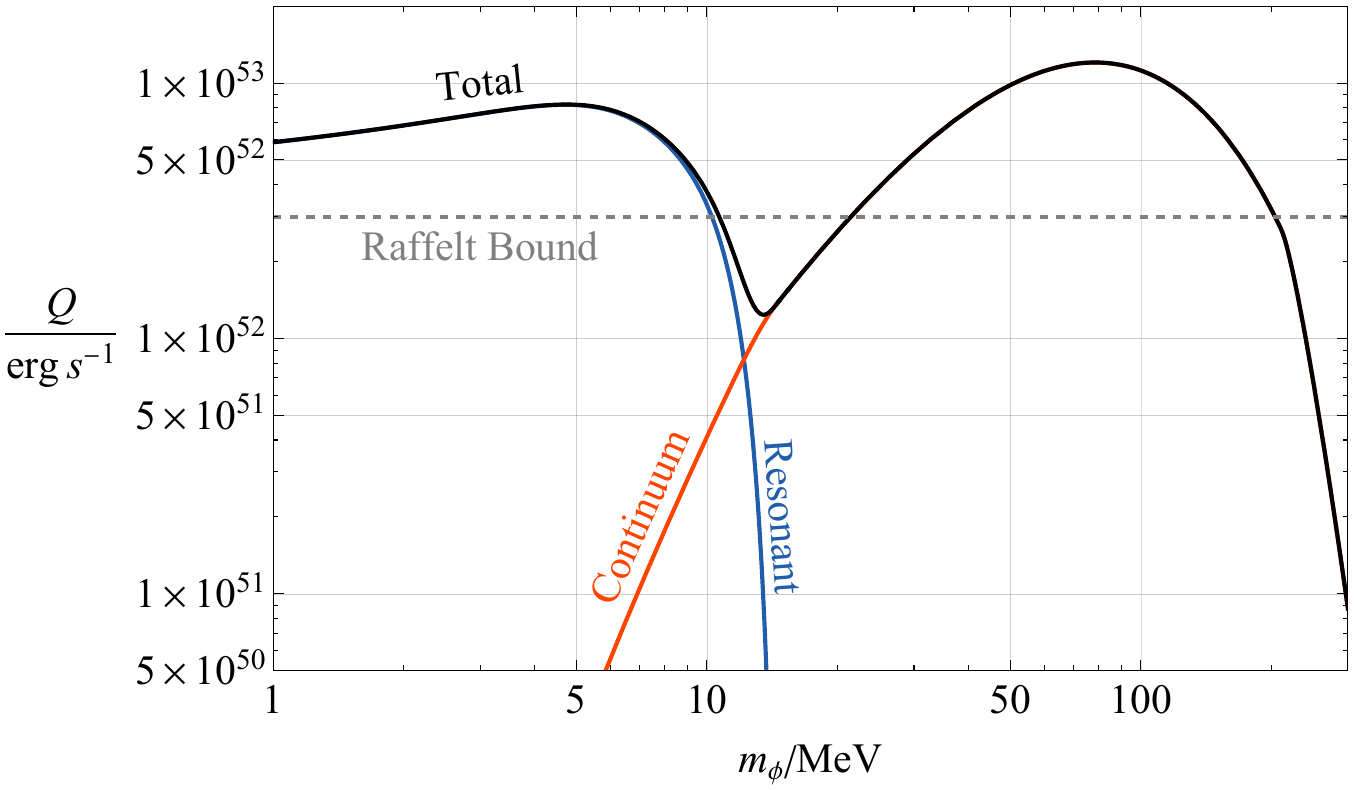}
    \caption{The power $Q$ emitted from a protoneutron star into scalars as a function of the scalar mass $m_\phi$, showing the contributions from resonant and continuum (bremsstrahlung) production. The scalars are assumed to couple to the Standard Model via mixing with the Higgs with mixing angle $\sin{\theta} = 3\times 10^{-6}$. These results are obtained using the $20~M_\odot$ supernova progenitor. The effects of reabsorption and decay of the scalar particles as they travel out of the supernova core, discussed in Section~\ref{sec: Trapping and decay}, are included although these have only a minor effect for the small value of $\sin\theta$ used. We also indicate the ``Raffelt Bound", above which the observed neutrino signal from SN~1987A is likely to have been altered, see Section~\ref{rfc} for a detailed discussion.}
    \label{fig:Luminosity vs Mass}
\end{figure}

The first term of Eq.~\eqref{eqn: diagram contributions to self energy} can be evaluated using the cutting rules, and contains contributions from nucleon bremsstrahlung, semi-Compton scattering, and pair annihilation, which will be discussed in Section~\ref{sec: continuum production}. We refer to these production mechanisms collectively as ``continuum production". The second term of Eq.~\eqref{eqn: diagram contributions to self energy} corresponds to mixing between the scalar and the longitudinal photon and features a resonance when the internal photon propagator goes on-shell, which is possible when $m_\phi$ is smaller than the plasma frequency. Hence we refer to this production mechanism as ``resonant production", the details of which will be analysed in Section~\ref{sec: resonant production}. As an advance summary, in Figure~\ref{fig:Luminosity vs Mass} we show the contributions of continuum and resonant production to the power output from a supernova into a scalar with Higgs-mixed couplings and $\sin\theta=3\times 10^{-6}$ fixed, as a function of the scalar's mass. For $m_\phi\lesssim 10~{\rm MeV}$  resonant production is possible and dominates over continuum production, which is strongly suppressed at small $m_\phi$. At larger $m_\phi$ continuum production is efficient, until $m_\phi\simeq 200~{\rm MeV}$ above which production is strongly kinematically suppressed.

\begin{figure}
    \centering
    \includegraphics[width=0.65\linewidth]{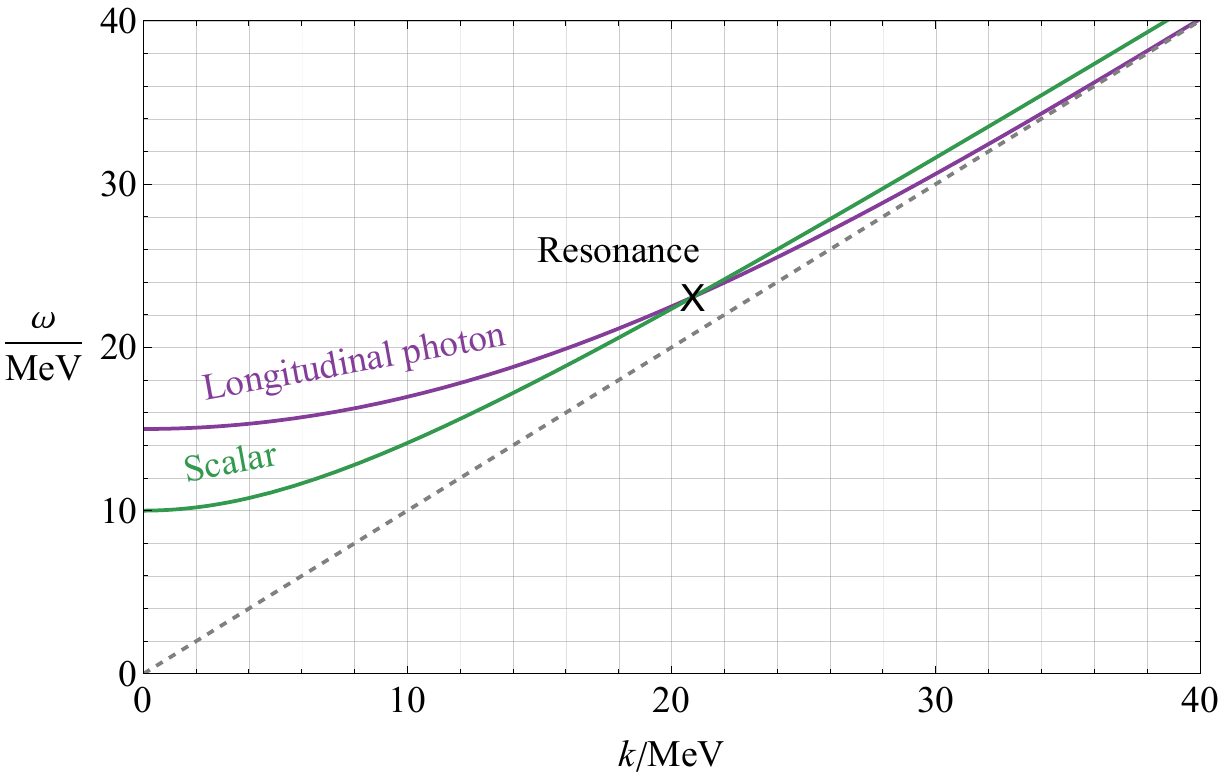}
    \caption{The dispersion relation for the longitudinal photon in a medium typical of a supernova core, with a plasma frequency of $15~{\rm MeV}$ and ultra-relativistic degenerate electrons. Also plotted is the free dispersion relation of a scalar with mass $10~{\rm MeV}$. Resonant production of the scalar occurs at the frequency where these curves cross.}
    \label{fig:dispersion crossing}
\end{figure}

\subsection{Resonant production}\label{sec: resonant production}
We first consider the second term of Eq.~\eqref{eqn: diagram contributions to self energy}. We expect there to be a resonance when the internal resummed thermal photon propagator goes on-shell. Physically this corresponds to the point where the scalar's dispersion relation crosses the photon's in-medium dispersion relation, since the diagram is evaluated for an on-shell scalar 4-momentum. The crossing is illustrated in Figure~\ref{fig:dispersion crossing} for typical dispersion relations. We will see that this physical expectation matches the result of our calculation.

In order to evaluate the second term of Eq.~\eqref{eqn: diagram contributions to self energy}, we must find the resummed $AR$ photon propagator $D_\gamma^{AR}$ and also the $RA$ mixing self-energy between the photon and the scalar $\Pi^{RA}_{\gamma\phi}$. The propagator $D_\gamma^{AR}$ gets contributions from the $RA$ photon self-energy $\Pi_{\gamma\gamma}^{RA}$ of the form 
\begin{equation}\label{eqn: resummed RA propagator}
\begin{tikzpicture}[baseline={(current bounding box.south)}]
  \begin{feynman}
    \coordinate (i1) at (-0.7, 0);
    \coordinate (o1) at (0.7, 0);

    \draw[photon] (i1) -- (o1);
    \draw[photon] (i1) ++(0,0.07) -- ++(1.4,0);

    \node at (-0.7,0.4) {\small A};
    \node at (0.7,0.4) {\small R};
    
  \end{feynman}
\end{tikzpicture}
=
\begin{tikzpicture}[baseline={(current bounding box.south)}]
  \begin{feynman}

    \coordinate (i1) at (-0.7, 0);
    \coordinate (o1) at (0.7, 0);

    \diagram* {
      (i1) -- [photon] (o1)
    };
    \node at (-0.7,0.4) {\small A};
    \node at (0.7,0.4) {\small R};
    
  \end{feynman}
\end{tikzpicture}
+
\begin{tikzpicture}[baseline={([yshift=-2.5pt] current bounding box.center)}]
  \begin{feynman}

    \coordinate (i1) at (-1.5, 0);
    \coordinate (o1) at (1.5, 0);
    \coordinate (b1) at (0, 0);
    
    \diagram* {
      (i1) -- [photon] (b1),
      (b1) -- [photon] (o1),
    };

    \filldraw[fill=gray!30] (0, 0) circle (0.6);

    \node at (b1) {\small $\Pi_{\gamma\gamma}^{RA}$};
    \node at (-1.5,0.4) {\small A};
    \node at (-0.8,0.4) {\small R};
    \node at (0.8,0.4) {\small A};
    \node at (1.5,0.4) {\small R};
    
  \end{feynman}
\end{tikzpicture}
\;
+\;
\begin{tikzpicture}[baseline={([yshift=-2.5pt] current bounding box.center)}]
  \begin{feynman}

    \coordinate (i2) at (-2.55, 0);
    \coordinate (b1) at (-1.05, 0);
    \coordinate (b2) at (1.05, 0);
    \coordinate (o2) at (2.55, 0);

    \diagram* {
      (i2) -- [photon] (b1),
      (b1) -- [photon] (b2),
      (b2) -- [photon] (o2),
    };

    \filldraw[fill=gray!30] (-1.05, 0) circle (0.6);
    \filldraw[fill=gray!30] (1.05, 0) circle (0.6);

    \node at (b1) {\small $\Pi_{\gamma \gamma}^{RA}$};
    \node at (b2) {\small $\Pi_{\gamma \gamma}^{RA}$};

    \node at (-2.55,0.4) {\small A};
    \node at (-1.85,0.4) {\small R};
    \node at (-0.25,0.4) {\small A};
    \node at (0.25,0.4) {\small R};
    \node at (1.85,0.4) {\small A};
    \node at (2.55,0.4) {\small R};
    
  \end{feynman}
\end{tikzpicture}
\;
+\dots~.
\end{equation}
Evidently our choice of an off-diagonal basis has made this resummation straightforward, whereas it would be much more complicated in the 1/2 basis.

We proceed by splitting up the Lorentz structure of the self-energy into transverse and longitudinal parts
\begin{equation}
    \Pi_{\gamma,\mu\nu}^{RA} = (\epsilon_\mu^+\epsilon_\nu^{+*}+\epsilon_\mu^-\epsilon_\nu^{-*})\Pi_{T}^{RA}+\epsilon_\mu^L\epsilon_\nu^{L}\Pi_{L}^{RA} ~.
\end{equation}
The physical condition that the scalar can only mix with the longitudinal mode of the photon translates to the condition 
\begin{equation} \label{eq:transverse_vanishes}
    \epsilon^{\pm}_\mu \Pi_{\phi \gamma}^{\mu,RA}=0~,
\end{equation}
which simply states that the mixing vanishes when the external photon is transverse. Eq.~\eqref{eqn: Mixing self energy} in Appendix~\ref{appendix: Relativistic deg mixing} shows that Eq.~\eqref{eq:transverse_vanishes} is indeed satisfied by the one-loop expression for the mixing that we use.
Consequently we can drop the transverse parts of the photon self-energy, because we will always be contracting the photon propagator with the mixing self-energy. This observation, along with the Ward identity $k_\mu \Pi_{\phi \gamma}^{\mu,RA}=0$, allows the resummation to be performed, leading to the compact expression for the ``mixing" contribution to the self-energy:  
\begin{equation} \label{eq:mixing_simplified}
    (-i \Pi_{\phi \gamma,\mu}^{RA})D^{AR,\mu\nu}_{\gamma}(-i \Pi_{\gamma\phi,\nu}^{RA}) = i\frac{\Pi^{RA}_{\phi L}\Pi^{RA}_{L\phi}}{\Pi_L^{RA}-K^2}~.
\end{equation}
Here we have defined $\Pi^{RA}_{\phi L} = \epsilon^L_\mu \Pi_{\phi\gamma}^{\mu,RA}$, and as before everything is evaluated with an on-shell scalar 4-momentum $K^2 = \omega^2-\boldsymbol{k}^2 = m_\phi^2$. Eq.~\eqref{eq:mixing_simplified} can be simplified further by noting that $\Pi^{RA}_{L\phi}(K) = \Pi^{AR}_{\phi L}(-K)=\Pi^{RA}_{\phi L}(K)$, where the first equality follows from the structure of the diagram and the second follows from the fact that the retarded propagator and the advanced propagator are related by changing the sign of the four-momentum (the second equality can also be derived from the spectral representation of the self-energies as shown in Ref.~\cite{Gelis:1997}). 
We then arrive at the production rate through resonant emission\footnote{In our resummation we have dropped the $i\epsilon$ prescription from the advanced/retarded propagators that appear, meaning our expression for $\Gamma_{\rm prod}^{\rm res}$ is not well defined if $K^2 = \Pi^{RA}_L(K)$. However, this never occurs because the self-energy has a small imaginary part corresponding to the width of the resonance.}
\begin{equation}
     \frac{dN_{\text{prod}}}{dV dt} = \int\frac{d^3\boldsymbol{k}}{(2\pi)^3}\Gamma_{\text{prod}}^{\text{res}}(\omega) =\int\frac{d^3\boldsymbol{k}}{(2\pi)^3} \frac{n_B(\omega)}{\omega}\text{Im}\left[\frac{(\Pi_{\phi L}^{RA})^2}{\Pi_L^{RA}-m_\phi^2}\right] \, ,
\end{equation}
which agrees with previous work, where it was derived by considering the diagonalization of the mixing matrix between the photon and the scalar \cite{Hardy:2017}. Our method emphasises how this expression arises in the framework of thermal field theory and avoids subtle analytic continuations that appear in the imaginary-time formalism, providing a complementary viewpoint.

By expanding  $\Pi_{\phi L}^{RA}$ and $\Pi_L^{RA}$ in their real and imaginary parts, and assuming that $\text{Im}[\Pi_{\phi L}^{RA}]\ll\text{Re}[\Pi_{\phi L}^{RA}]$ (see Appendix~\ref{appendix: Relativistic deg mixing}), we can write the resonant production rate in a form that makes the physical properties of the resonance clear:
\begin{equation}\label{eqn: resonant production pre delta function}
    \frac{dN_{\text{prod}}}{dV dt} = \int\frac{d^3\boldsymbol{k}}{(2\pi)^3}\frac{n_B(\omega)}{\omega}\left(\frac{\omega}{m_\phi}\text{Re}[\Pi_{\phi L}^{RA}]\right)^2\frac{\sigma_L \omega}{(\sigma_L \omega)^2+(\omega^2-\omega_L^2)^2} \, ,
\end{equation}
where we define $\sigma_L$ and $\omega_L$ such that $\text{Im}[\Pi^{RA}_L]= \frac{K^2}{\omega^2}\omega \sigma_L$ and $\text{Re}[\Pi^{RA}_L]=  \frac{K^2}{\omega^2} \omega_L^2$.

Evidently, the resonance occurs at $\omega = \omega_L$. This is equivalent to the condition $K^2 - \text{Re}[\Pi^{RA}_L] = 0$, which is exactly the photon's mass-shell condition matching the physical interpretation given at the beginning of this section. Due to the shape of the longitudinal photon's dispersion relation, there is always an energy and momentum where the resonance condition is met provided $m_\phi<\omega_p$, where the plasma frequency $\omega_p$ is defined as the longitudinal photon's energy at zero momentum. This means that resonant production can occur at every point within the supernova core, out to some maximum radius where the decreasing temperature and density cause the plasma frequency to drop below the scalar's mass. We refer to this inner region as the ``resonance region" for a given $m_\phi$. Because of the form of the supernova profiles, for a particular mass and energy there is at most one value of the radius where the resonance condition is met, and we refer to this as the ``resonant shell". 

The width of the resonance is set by $\text{Im}[\Pi^{RA}_L]$, via $\sigma_L$, which encodes the lifetime of the longitudinal photon. Given that $\sigma_L\ll\omega_L$,\footnote{This is supported by the fact that $\text{Im}[\Pi^{RA}_L]$ has no contribution at leading order, because, despite the photon's effective in-medium mass, decays into electrons are forbidden by the electron also acquiring a thermal mass~\cite{Braaten:1993jw}.} the longitudinal photon is a narrow resonance and the final factor in Eq.~\eqref{eqn: resonant production pre delta function} can be approximated by a delta function. Performing the integral over the scalar's momentum leaves
\begin{equation} \label{eq:prod_res}
    \frac{dN_{\text{prod}}}{dV dt} = \frac{1}{4\pi}\frac{k_*}{\omega_*}n_B(\omega_*)\left(\frac{\omega_*}{m_\phi}\text{Re}\left[\Pi_{\phi L}^{RA}(\omega_*)\right]\right)^2\left[1-\frac{\partial \omega_L^2}{\partial \omega^2}\right]^{-1}_{\omega = \omega_*},
\end{equation}
where $(\omega_*,\boldsymbol{k}_*)$ is the 4-momentum at the resonant frequency, and $k_* = |\boldsymbol{k_*}|$. 
The final term in Eq.~\eqref{eq:prod_res} arises 
from the delta function 
and is closely related to the ``residue factor" in thermal field theory \cite{Le_Bellac:1996}, which characterises the fact that collective excitations such as the longitudinal photon may couple with a different strength compared to their vacuum counterparts. 

In Appendix~\ref{appendix: Relativistic deg mixing} we show that the leading, one-loop, contributions to $\text{Re}[\Pi_{\phi L}^{RA}]$ from protons and electrons are, respectively,
\begin{equation}
\begin{aligned} \label{eq:mixing_both}
    \text{Re}[\Pi_{\phi L}^{RA}(\omega_*)]_{\text{protons}} \; = \;  \frac{g_P e m_{\phi} n_P k_*}{\omega_*^2 m_P}\, ,  \\
       \text{Re}[\Pi_{\phi L}^{RA}(\omega_*)]_\text{electrons} \; = \; \frac{g_e m_\phi m_e k_*}{e\mu_e}\, ,
\end{aligned}
\end{equation}
where $g_{P,e}$ are the scalar's couplings to protons and electrons, $n_P$ is the number density of protons, and both the scalar and the longitudinal photon are put on-shell. In Eq.~\eqref{eq:mixing_both}, we make the approximation that protons are non-relativistic (note that this expression holds regardless of their degeneracy) and assume that the electrons are relativistic and degenerate with chemical potential $\mu_e \gg T$. Despite the electrons being the dominant contribution to the photon self-energy \cite{Raffelt:1996wa, Kopf_1998}, for the case of the Higgs-mixed scalar the mixing self-energy in supernova conditions is dominated by the protons. This is partly because, as discussed in Ref. \cite{DeRocco:2022rze}, scalar-photon mixing requires a chirality flip and is therefore suppressed for electrons in the ultra-relativistic limit. 
Substituting $\text{Re}[\Pi_{\phi L}^{RA}(\omega_*)]_{\text{protons}}$ into Eq.~\eqref{eq:prod_res} the rate of energy loss due to proton-mediated mixing is 
\begin{equation}
    \frac{dQ}{dVdt}_{\;\text{protons}} = 2 \alpha g_P^2\left(\frac{n_P^2}{m_P^2}\right)\frac{1}{e^{\omega_*/T}-1}\omega_* v_*^5\left(2+\frac{m_\phi^2-3\omega_p^2}{\omega_*^2}\right)^{-1},
\end{equation}
where $v_* = \sqrt{1-m_\phi^2/\omega_*^2}$. 
For the case of a leptophilic scalar we will need the energy loss due to electron-mediated mixing, which is similarly immediately obtained from Eqs.~\eqref{eq:prod_res} and~\eqref{eq:mixing_both}.

Since the photon's plasma frequency is dominated by the contribution from relativistic electrons, the plasma frequency is given by \cite{Raffelt:1996wa}
\begin{equation} \label{eqn: plasma frequency}
    \omega_p^2 = \frac{4\alpha}{3\pi}\left(\mu_e^2+\frac{1}{3}\pi^2 T^2\right).
\end{equation}
To find the resonant frequency $\omega_*$, one must solve the two dispersion relations $\omega^2-\boldsymbol{k}^2=m_\phi^2$ and $\omega^2-\boldsymbol{k}^2 = \text{Re}\left[\Pi_L^{RA}(\omega,\boldsymbol{k})\right]$ simultaneously. There is a well-known analytic approximation for $\text{Re}\left[\Pi_L^{RA}\right]$, which can be conveniently written as \cite{Braaten:1993jw,Raffelt:1996wa}
\begin{equation}\label{eqn: photon self energy}
    \text{Re}\left[\Pi_L^{RA}\right] = \frac{K^2}{\omega^2}\omega_p^2\left(1+H(v_0^2k^2/\omega^2)\right),~~{\rm where}~~     H(x)=\frac{3}{2x^{3/2}}\log\left(\frac{1+\sqrt{x}}{1-\sqrt{x}}\right)-1-\frac{3}{x}~.
\end{equation}
Here $v_0$ is the electron velocity that dominates the phase space integration. In our case, the photon self-energy is dominated by ultra-relativistic electrons so $v_0 \simeq 1$.

\subsection{Continuum production}\label{sec: continuum production}

\begin{figure}
    \centering
    \resizebox{0.32\columnwidth}{!}{
    \begin{tikzpicture}

\draw[dashed, thick] (-3,0) -- (-1.5,0);
\draw[dashed, thick] (3,0) -- (1.5,0);

\node at (-1.8,0.3) {$R$};
\node at (1.8,0.3) {$A$};

\draw[thick, postaction={decorate}, decoration={
    markings, 
    mark=at position 0.125 with {\arrow{Straight Barb}},
    mark=at position 0.375 with {\arrow{Straight Barb}},
    mark=at position 0.625 with {\arrow{Straight Barb}},
    mark=at position 0.875 with {\arrow{Straight Barb}}
}] 
(0,0) circle (1.5);

\node at (-3.3,0) {$\phi$};
\node at (3.3,0) {$\phi$};

\node at (1.4,-1.4) {$e$};

\end{tikzpicture}
    }
    \hfill
    \resizebox{0.32\columnwidth}{!}{
    \begin{tikzpicture}

\draw[dashed, thick] (-3,0) -- (-1.5,0);
\draw[dashed, thick] (3,0) -- (1.5,0);

\node at (-1.8,0.3) {$R$};
\node at (1.8,0.3) {$A$};

\draw[thick, postaction={decorate}, decoration={
    markings, 
    mark=at position 0.125 with {\arrow{Straight Barb}},
    mark=at position 0.375 with {\arrow{Straight Barb}},
    mark=at position 0.625 with {\arrow{Straight Barb}},
    mark=at position 0.875 with {\arrow{Straight Barb}}
}] 
(0,0) circle (1.5);

\draw[thick, decorate, decoration={snake, amplitude=2pt, segment length=8pt}] (0,1.5) -- (0,-1.5);

\node at (-3.3,0) {$\phi$};
\node at (3.3,0) {$\phi$};

\node at (0.3,0) {$\gamma$};

\node at (1.7,-1.2) {$p$/$e$};

\end{tikzpicture}
    }
    \hfill
    \resizebox{0.32\columnwidth}{!}{
    \begin{tikzpicture}

    \draw[dashed, thick] (-3,0) -- (-1.5,0);
    \draw[dashed, thick] (3,0) -- (1.5,0);
    
    \node at (-1.8,0.3) {$R$};
    \node at (1.8,0.3) {$A$};

    \draw[thick, postaction={decorate}, decoration={markings, 
        mark=at position 0.125 with {\arrow{Straight Barb}}, 
        mark=at position 0.375 with {\arrow{Straight Barb}}, 
        mark=at position 0.625 with {\arrow{Straight Barb}}, 
        mark=at position 0.875 with {\arrow{Straight Barb}}}] 
        (0,0) circle (1.5);
    
    \draw[thick, postaction={decorate}, decoration={markings, 
        mark=at position 0 with {\arrow{Straight Barb}}, 
        mark=at position 0.5 with {\arrow{Straight Barb}}}] 
        (0,0) circle (0.5);
    
    \draw[dotted, thick] (0,-1.5) -- (0,-0.5);
    \draw[dotted, thick] (0,0.5) -- (0,1.5);

    \node at (0.2,-1) {$\pi$};
    \node at (0.2,1) {$\pi$};

    \node at (1.4,-1.2) {$N$};
    \node at (0.8,0) {$N$};

    \node at (-3.3,0) {$\phi$};
    \node at (3.3,0) {$\phi$};

\end{tikzpicture}
}
    \caption{Three of the lowest order diagrams contributing to $\text{Im}\left[\Pi^{RA}_{\phi\phi}\right]$. When evaluated using the cutting rules, the first corresponds to pair annihilation $e^+e^-\rightarrow\phi$ (although there is no phase space for this process), the second contributes to ``semi-Compton'' scattering, e.g. $e^- \gamma \rightarrow e^- \phi$, and also higher-order annihilation $e^+e^-\rightarrow\phi\gamma$, and the third contributes to bremsstrahlung as calculated in the one-pion-exchange approximation $NN \rightarrow NN\phi$.}
    \label{fig:Lowest order loops}
\end{figure}
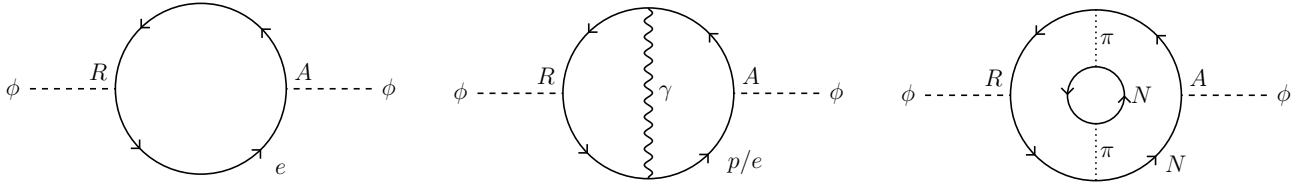

We now analyse the production of scalars due to the first term of Eq.~\eqref{eqn: diagram contributions to self energy}, corresponding to the 1PI self-energy $\text{Im}\left[\Pi^{RA}_{\phi\phi}\right]$.
Three of the lowest order diagrams that we consider are shown in Figure~\ref{fig:Lowest order loops}. When calculated using the cutting rules these correspond to familiar physical processes, and the rules automatically account for the usual Pauli-blocking or Bose-enhancement factors. In particular, the diagrams of Figure~\ref{fig:Lowest order loops} give contributions to pair annihilation, semi-Compton scattering, and bremsstrahlung respectively. Which of these processes dominates of course depends on $\phi$'s couplings to the SM. 

In the case of a Higgs-mixed scalar we expect the dominant production mechanism to be nucleon-nucleon bremsstrahlung, due to the large nucleon number density as well as the fact that it is mediated by the strong force. Meanwhile, any process in which the scalar couples directly to electrons will be suppressed due to the small Higgs-electron coupling as well as the small electromagnetic coupling. Although pion semi-Compton production $\pi N\rightarrow \phi N$ is also mediated by the strong interaction, this will be proportional to the pion number density. The pion number density is rather dependent on the details of the nuclear equation of state inside the protoneutron star, which has large uncertainties \cite{Migdal:1990vm} but is likely to be small relative to the nucleon number densities \cite{Lattimer:1991nc, Janka:2006fh, Fore:2019wib}. Ref.~\cite{Carenza:2020cis} finds that $\pi^{-}$ semi-Compton is relevant for the case of axions, however we leave the inclusion of this process for scalars to future work.

We thus need to calculate the bremsstrahlung rate
\begin{equation}
        Q_{\text{br}} = \int\frac{d^3\boldsymbol{k}}{(2\pi)^3}\omega\Gamma_{\text{prod}}^{\text{br}}(\omega)~,
\end{equation}
where
\begin{align}\label{eqn: brem rate pre integratoin}
    \Gamma_{\text{prod}}^{\text{br}}(\omega)=\frac{1}{2\omega}\int \prod_{i=1}^4\frac{d^3\boldsymbol{p_i}}{(2\pi)^3}\frac{\mathcal{S}}{2E_i} f_1 f_2 (1-f_3)(1-f_4)(2\pi)^4\delta^4\left(\sum_i P_i^\mu-K^\mu\right)\sum_{\text{spins}}|\mathcal{M}_{NN\phi}|^2~.
\end{align}
Here $\mathcal{S}$ is a symmetry factor (1 for neutron-proton bremsstrahlung, and 1/4 for proton-proton and neutron-neutron bremsstrahlung), $\sum_i P_i^\mu = P_1^\mu+P_2^\mu-P_3^\mu-P_4^\mu$, and $f_i$ are the Fermi distributions for the different momenta ($P_{1,2}$ and $P_{3,4}$ are the momenta of the incoming and outgoing nucleons respectively and $K$ is the momentum of the emitted scalar). The major obstruction to calculating this rate is evaluating the matrix element $|\mathcal{M}_{NN\phi}|^2$, which is a challenging nuclear physics problem. The matrix element can be calculated in the one-pion-exchange approximation, however it has been shown that such an approximation systematically overestimates the nucleon-nucleon scattering cross section by almost an order of magnitude \cite{Rrapaj:2015wgs}. This can be addressed in a number of ways, for example by changing the pion propagator to include the exchange of a $\rho$-meson, which mimics the effect of two pion exchange diagrams \cite{Carenza:2019pxu}. We choose to use a soft approximation in order to relate the matrix element to $2\rightarrow 2$ nucleon scattering cross sections, which are measured in experiments, thereby evading the uncertainties in the nuclear physics calculations \cite{Rrapaj:2015wgs}. 
As we will see, this also simplifies the phase space integration significantly. A detailed comparison of the OPE approximation and the soft approximation, along with a discussion of their respective uncertainties, is provided in Appendix~\ref{appendix: OPE comparison}.

Assuming non-relativistic nucleons and expanding in the small parameters $m_\phi^2/(2m_N \omega)$ and $\slashed{K}/(2m_N)$,
a straightforward calculation shows that the matrix element $\mathcal{M}_{NN\phi}$ is related to the $2\rightarrow 2$ scattering matrix element $\mathcal{M}_{NN}$ by
\begin{equation}\label{eqn: soft matrix element}
    |\mathcal{M}_{NN\phi}|^2 = 4g_N^2\frac{m_\phi^4}{m_N^2}\frac{1}{\omega^4}|\mathcal{M}_{NN}|^2~,
\end{equation}
where, as before, $\omega$ denotes the scalar's energy, and $g_N$ is the coupling between the scalar and the nucleon.\footnote{By expanding to the next order in the nucleon momenta, we find a reduction to the matrix element by a factor of $\frac{3}{4}$, which is comparable to other uncertainties.} We make the approximation that the proton and neutron masses and couplings to $\phi$ are equal.  
As a result, the spin-summed matrix element for $\phi$ production is given in terms of the differential cross section for nucleon-nucleon scattering by
\begin{equation}\label{eqn: M in terms of cross section}
    \sum_{\text{spins}}|\mathcal{M}_{NN\phi}|^2 = 1024\pi^2g_N^2\frac{m_\phi^4}{m_N^2}\frac{1}{\omega^4}E_{\text{CM}}^2\left(\frac{d\sigma}{d\Omega}\right)_{\text{CM}}~,
\end{equation}
where $E_\text{CM}\approx 2m_N$ is the total energy of the nucleons in the centre of mass frame.

Turning to the phase space integrals, since we are working in the soft approximation we drop the scalar's 4-momentum from the delta function responsible for energy/momentum conservation. Physically this corresponds to assuming that the nucleons do not recoil after emitting a scalar and means that the 
phase space integration is the same for both production and inverse bremsstrahlung absorption, so we treat them both together. Although the previous literature considering scalars used a Boltzmann distribution for $f_i$, we find that for the supernova density and temperature profiles that we consider the nucleons can be moderately degenerate in the core with a degeneracy parameter $\eta_i = \mu_i/T \lesssim 6$ (see Appendix~\ref{appendix: sn profiles} for further details). Consequently, we use the full Fermi-Dirac distribution, allowing for Pauli-blocking. 
Due to the simple structure of the matrix element, the production and absorption rates by bremsstrahlung $\Gamma_{\text{prod/abs}}^{\text{br}}(\omega)$ can be written in terms of the total nucleon-nucleon scattering cross section $\sigma(E_K^\text{CM})$ integrated over the centre of mass kinetic energy $E_K^\text{CM}$:
\begin{equation}\label{eqn: brem rate}
    \Gamma_{\text{prod/abs}}^{\text{br}}(\omega)=\frac{8\mathcal{S}g_N^2 n_1 n_2 m_\phi^4}{\pi m_N^2}\sqrt{\frac{\pi T}{m_N}}\frac{1}{\omega^5}\int_{x_{\text{min}}}^{\infty} dx \, x \, \sigma(x T) \, \Bar{\Sigma}(x,\eta_1,\eta_2)~,
\end{equation}
where we have introduced the dimensionless quantity $x = E_K^\text{CM}/T$, $n_1$ and $n_2$ are the number densities of the nucleons involved, and $\Bar{\Sigma}(x,\eta_1,\eta_2)$ is defined in terms of an integral over a complicated function given in Appendix~\ref{appendix: degeneracy bremsstrahlung}. 
In the limit of zero degeneracy $\Bar{\Sigma}(x,\eta_1,\eta_2)\rightarrow e^{-x}$, reproducing a result similar to the dark photon production rate presented in Ref. \cite{Chang:2016ntp}. For production, we introduce a lower bound $x_{\min} = \omega/T$ by hand on the integral in Eq.~\eqref{eqn: brem rate}. This imposes that the collision energy is large enough to produce the scalar, compensating for removing the scalar's 4-momentum from the delta function. 
For absorption there is no such constraint so in this case we take $x_{\text{min}}=0$. From here it is straightforward to calculate the production rate by using a parameterization of the nuclear scattering cross sections, for example as found in Ref. \cite{Norbury2013NucleonNucleonTC}, and summing over  neutron-neutron, neutron-proton, and proton-proton collisions.

On general grounds, the production and absorption rate from the thermal bath should be related by detailed balance $\Gamma_{\text{prod}}(\omega) = e^{-\omega/T}\Gamma_{\text{abs}}(\omega)$. 
However, we find that in the degenerate core the production rate calculated in Eq.~\eqref{eqn: brem rate} for typical values of $\omega$ is two to three orders of magnitude larger than that calculated using detailed balance. The reason for this discrepancy is the interplay between the ``no nucleon recoil" approximation and Pauli-blocking. Consider the unphysical but conceptually simpler case of highly degenerate nucleons. For a nuclear collision occurring near the Fermi surface, emission of a scalar would decrease the energy of the nucleons. If this would cause the energy of the nucleons to go below the Fermi surface then the process is Pauli-blocked. However, in the ``no nucleon recoil" approximation such blocking is missed, and the emission would be allowed. Therefore our soft approximation overestimates emission. 
By similar logic, $\Gamma_{\text{abs}}(\omega)$ is underestimated (because the no-recoil approximation misses the possibility that a nucleon beneath the Fermi surface can be lifted out by the energy gained by absorbing a scalar, and therefore overestimates the impact of Pauli-blocking). 
Unfortunately, calculating the phase space integrals accounting for the scalar's momentum is too computationally expensive  to be viable when using the full Fermi-Dirac distributions. 
Therefore, in order to obtain conservative limits, we choose to calculate $\Gamma_{\text{prod}}(\omega)$ using detailed balance and $\Gamma_{\text{abs}}(\omega)$ evaluated in the soft approximation with no nucleon recoil, which underestimates rather than overestimates the production rate. We provide numerical evidence for this in Appendix~\ref{appendix: no nucleon recoil}.

\section{Reabsorption and decay}\label{sec: Trapping and decay}
Once the scalars have been produced they still have to escape the protoneutron star in order to contribute to the anomalous energy loss relevant for the Raffelt bound. Two things may prevent this escape: reabsorption, and decay into Standard Model degrees of freedom (which then equilibrate with the plasma). In the limit of small coupling both of these processes can be neglected leading to the free-streaming regime, however for larger couplings they can be important leading to the trapping regime. The trapping regime introduces a significant dependence on the details of the particular BSM theory that includes the new scalar, due to the various decay channels that may be available in different theories. For example, if the scalar is a mediator to a dark sector that does not directly interact with the SM then it might quickly decay to hidden sector states, which could then free-stream out of the star (reabsorption and self-interaction of these states may change this simple picture~\cite{Sokolov:2019cbs,Fiorillo:2024upk}). In this section we consider only minimal models in which the new scalar is the only light degree of freedom, but our calculations can easily be adapted to more complicated theories.

The impact of decays and reabsorption can be incorporated by multiplying the emission rate per unit volume and energy by an ``attenuation factor". This represents the probability, averaged over the solid angle of emission, of a produced scalar propagating out of the neutrinosphere. In a full 3-dimensional model 
the attenuation factor is given by
\begin{equation}
    A(\boldsymbol{r},\omega) = \frac{1}{4\pi}\int d\Omega \, \exp\left[-\frac{1}{v}\int_{C(\boldsymbol{r},\Omega)}ds' \,\bigl(\Gamma_{\text{abs}}(\boldsymbol{r'},\omega)+\Gamma_{\text{dec}}(\boldsymbol{r'},\omega) \bigr) \right],
\end{equation}
where $C(\boldsymbol{r},\Omega)$ is the path from position $\boldsymbol{r}$ to the neutrinosphere, in the direction $\Omega$. 
Including this full expression in our results would be very computationally expensive, even with our assumption of spherical symmetry, due to the need to integrate over the position and energy of the emitted particles. 
Instead, we make the approximation of replacing the integral over the polar angle $\theta$ with the average of the two radial directions, $\theta = 0$ and $\theta = \pi$, which corresponds to assuming that half of the particles propagate radially inwards, and half propagate radially outwards (for more sophisticated approaches see e.g. Refs.~\cite{Caputo:2022rca,Lucente:2022wai}). 
To validate this approximation we have carried out the full angular integration for a model protoneutron star of constant density and temperature, in which case the half-in half-out approximation only differed from the full result by $\mathcal{O}(1)$ factors, which we expect to be comparable to the other uncertainties in the trapping regime. Importantly, recent developments in the theory of core collapse supernovae suggest that the cooling phase is essentially asymmetric~\cite{Nagakura:2019tmy, Burrows:2020qrp, Vartanyan:2021dmy}, so that the assumption of spherical symmetry is in reality substantially violated; this means that even performing the full angular integration in our spherically symmetric supernova models would not necessarily improve the accuracy of the inferred constraints.

The total power emitted from the protoneutron star  into scalars is then given by the compact expression
\begin{equation}\label{eqn: Total power emission}
    Q = \int_0^{R_{\nu}}\! dr \, 4\pi r^2\int\frac{d^3\boldsymbol{k}}{(2\pi)^3} \, \omega \Gamma_{\text{prod}}(r,\omega)A(r,\omega)~,
\end{equation}
where the radius of the neutrinosphere is $R_\nu\approx 30$~km in the reference supernova model that we use, with an insignificant variation of $\mathcal{O}(1)$~km depending on the progenitor mass and the neutrino species considered. In principle, all that remains is to evaluate this expression, taking into account all production, absorption and decay channels as well as finite temperature and density effects such as Pauli-blocking. As discussed at the end of Section~\ref{sec: continuum production}, in order to obtain conservative constraints we calculate $\Gamma_{\rm prod}$ from $\Gamma_{\rm abs}$ via detailed balance.\footnote{One could argue that for the most conservative bounds we should also overestimate the reabsorption rate by first calculating the production rate using Eq.~\eqref{eqn: brem rate} and then using detailed balance to relate it to the absorption rate. However, reabsorption will predominantly occur in the less degenerate outer regions where the complications of combining the no-recoil approximation and Pauli-blocking can be ignored.} 

Since the rates of resonant emission and absorption involve a delta function in energy and radius, we must split both the attenuation and the production into resonant and continuum parts. The attenuation factor then involves the exponent of a step function corresponding to whether the path passes through a resonant shell, which is schematically of the form $\exp\left[{-A\Theta(r_*(\omega)-r)}\right]$ where $r_*(\omega)$ is the position of the resonant shell for energy $\omega$. In practice, we find that $A$ is typically a large enough number that only a small fraction of scalars manage to pass through a resonant shell. 
As a result, for resonant production we need only consider scalars that are emitted outwards (otherwise they will be reabsorbed when they meet the resonant shell again) and these are only attenuated by inverse bremsstrahlung. Meanwhile, for continuum production there are three cases: First, if the scalar's energy is above the maximum resonant frequency anywhere inside the protoneutron star then absorption is always only via inverse bremsstrahlung. Second, for scalars with energy such that there is a resonant shell but which are produced outside this shell, only scalars emitted outwards are relevant. Finally, any scalars produced inside a resonant shell are always negligible compared to those produced directly by resonance (this is the case both in the free-streaming and trapping regimes).

Turning to decays, for Higgs-mixed scalars with masses in the range we are interested in, $0.1~{\rm MeV}\lesssim m_\phi \lesssim 300~{\rm MeV} $, decays to electrons, muons, and photons can be relevant. Because the decay to photons is loop-suppressed, we expect electrons and muons to be the dominant decay channels. In vacuum the rates for these decays are given by 
\begin{equation}
    \Gamma(\phi\rightarrow l^+ l^-) = \frac{m_\phi}{\omega}\left(\frac{m_l \sin{\theta}}{v_{\text{EW}}}\right)^2\frac{m_\phi}{8\pi}\left(1-\frac{4m_l^2}{m_\phi^2}\right)^{3/2}~,
\end{equation}
where $l$ is the lepton, either $e$ or $\mu$, and $m_l$ is the lepton's mass. 
While we expect the muon rate to be well-approximated by the vacuum rate, a number of complications arise for the electrons. First, due to the large electron number density in the core, the decays may be Pauli-blocked. 
Second, and much more challenging to account for, thermal effects modify the electrons' dispersion relation. Since the dispersion is no longer Lorentz invariant, one cannot simply go to the scalar's rest frame to solve for the energy and momentum of the produced electrons. Instead, conservation of energy and momentum become a pair of non-linear simultaneous equations that have to be solved numerically. In Appendix~\ref{appendix: In medium decay} we show that there is only a very small difference between the final constraints on a Higgs-mixed scalar obtained using the vacuum decay rate and those obtained making the rough approximation that the electron has an effective in-medium mass $\tilde{m}_e$  
and a simple dispersion relation $E = \sqrt{\boldsymbol{p}^2+\tilde{m}_e^2}$ (see Eqs.~\eqref{eq:me1} and~\eqref{eq:me2} in Appendix~\ref{appendix: In medium decay} for details), so for such a particle we simply use the vacuum rates. However, for the case of a leptophilic scalar we use the effective mass, which we find to have a minor but non-negligible impact.

\begin{figure}
    \centering
    \includegraphics[width=0.7\linewidth]{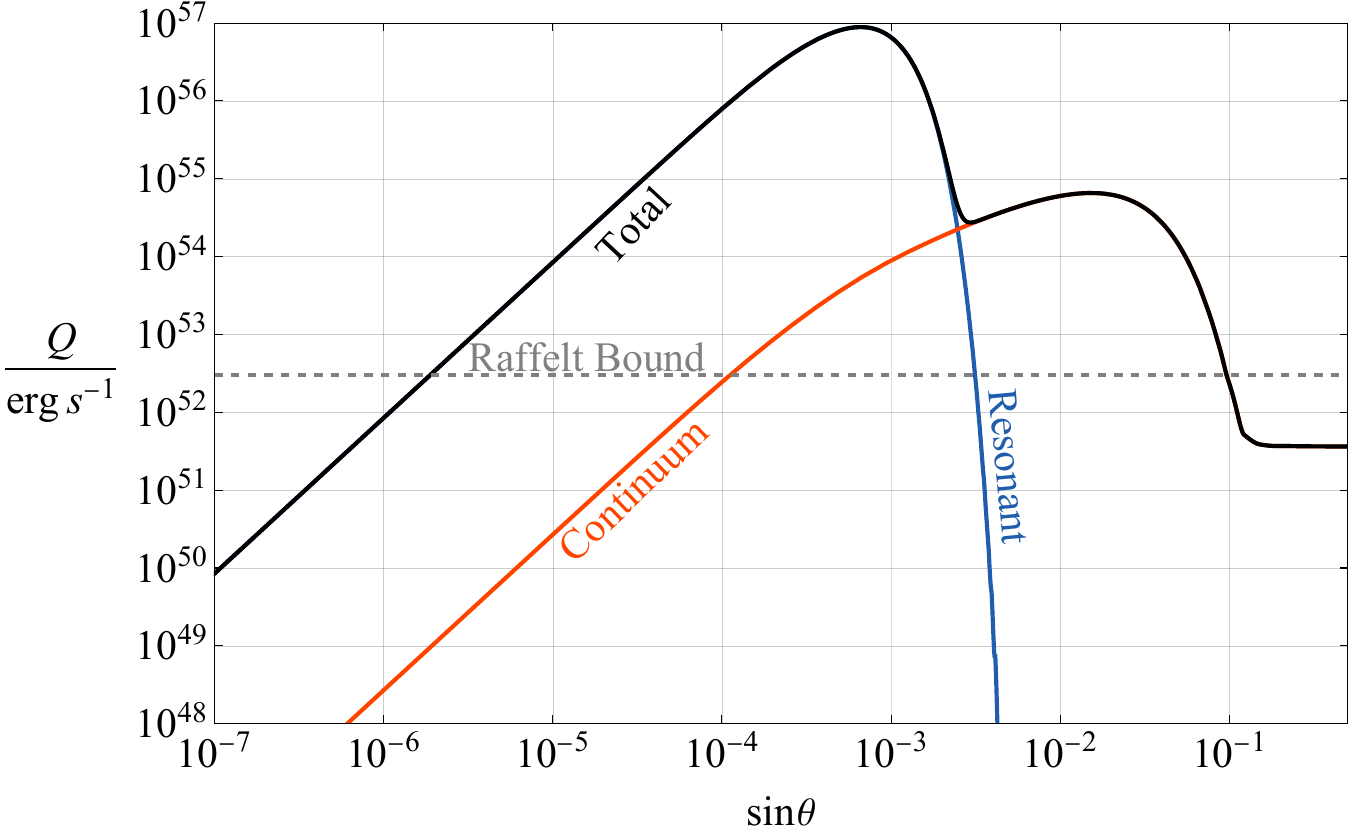}
    \caption{The total power emitted from a protoneutron star into Higgs-mixed scalars as a function of the mixing angle $\theta$, showing the contributions from resonant and continuum (bremsstrahlung) production. We fix $m_\phi = 3$ MeV and use the $20~M_\odot$ supernova progenitor. At $\sin\theta \gtrsim 10^{-3}$ attenuation by absorption and decays to leptons is important.
    }
    \label{fig:Luminosity vs coupling}
\end{figure}

To show the effects of absorption and decay, in Figure~\ref{fig:Luminosity vs coupling} we plot the total emitted power into a Higgs-mixed scalar as a function of the mixing angle $\sin\theta$ with $m_\phi=3~{\rm MeV}$ fixed. For $\sin\theta \lesssim 10^{-4}$ absorption and decays are negligible and the emitted energy simply increases with increasing coupling with resonant production dominating. For the value of $m_\phi$ plotted, resonant production occurs only inside a radius $r \approx 20~{\rm km}<R_\nu$ because outside this $\omega_p(r) < m_\phi$. If scalars produced at the outermost part of the resonance region are significantly attenuated then the resonant contribution is trapped. This happens at smaller $\sin\theta$ than for the continuum contribution, which can always be produced arbitrarily close to the neutrinosphere. As a result, continuum production dominates for $\sin\theta \gtrsim 3\times 10^{-3}$.  At very large $\sin\theta \approx 10^{-1}$ the continuum component is also trapped, and the plateau at large coupling\footnote{A similar plateau was observed in analogous studies for the case of other BSM particles~\cite{Lella:2023bfb, Lai:2024mse}.} corresponds to surface emission from the neutrinosphere such that $Q$ is approximately proportional to $4\pi R_{\nu}^2 T^4$ (however even in this regime the emission is from an extended region rather than being pure surface emission, which is consistent with previous literature \cite{Caputo:2021rux, Caputo:2022rca}). The total emitted power drops below the Raffelt bound at large $\sin\theta$ dominantly due to decays rather than reabsorption, and in fact this is the case for all $m_\phi>2m_e$.

\section{Results} \label{sec:results}
\subsection{Higgs-mixed scalars}

\begin{figure}
    \centering
    \includegraphics[width=0.9\linewidth]{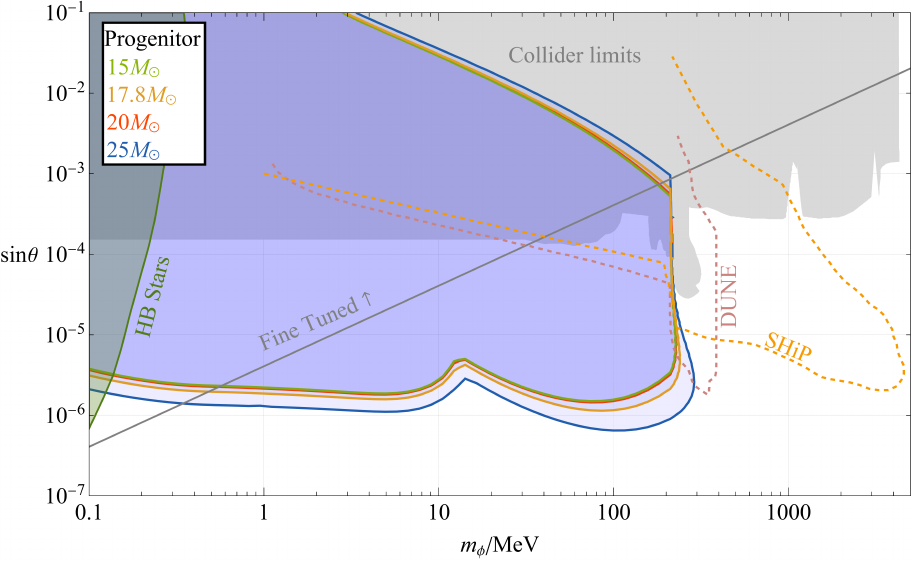}
    \caption{SN~1987A cooling constraints on a new scalar of mass $m_\phi$ that mixes with the Higgs, with mixing angle $\sin\theta$. We show results for four progenitors with masses around that of SN~1987A; the differences between these give an estimate of the associated uncertainty. We assume that the scalar does not decay to additional new degrees of freedom, and as a result for large mixing it can be reabsorbed or decay to Standard Model particles, which sets the upper and right hand edges of the constrained region. We also show constraints from collider searches, taken from Refs.~\cite{Balaji:2022noj,Lanfranchi:2243034}, and from horizontal branch stars, taken from Ref.~\cite{Knapen:2017xzo}, described in the main text.
    }
    \label{fig:Higgs-portal results}
\end{figure}

Combining the results of Sections~\ref{sec:production} and~\ref{sec: Trapping and decay}, we obtain the constraints on a Higgs-mixed scalar plotted in Figure~\ref{fig:Higgs-portal results}. To give an indication of the uncertainty from the progenitor, we show the bounds for the four discussed in Section~\ref{sec:supernovae}. The associated variability is largely determined by the difference in the PNS masses of the four progenitors, in particular we observe that the $15~M_{\odot}$ and $20~M_{\odot}$ progenitor models, which have similar PNS masses (see Section~\ref{sec:supernovae}), yield almost identical constraints, and the constraints become more stringent for the case of the progenitor models with larger PNS masses. Varying the bound on the allowed BSM energy loss $L_{\text{BSM}}$ over the range $[2\times 10^{52},4\times 10^{52}]~\text{erg}/\text{s}$ alters the results by less than the difference arising from the progenitors. Despite the neutron degeneracy parameter $\mu_n/T$  (where $\mu_n$ is the neutron chemical potential) reaching values of approximately $5$ in the supernova core, the only effect of degeneracy on the constraints is to slightly weaken the bound at large $m_\phi$ and small $\sin\theta$ (see Appendix~\ref{appendix: degeneracy bremsstrahlung} for further details). As discussed, the upper edge of the constrained region, set by trapping, depends on the details of the BSM theory considered. In the figure we assume that there are no additional hidden sector particles that the scalar can decay to. For such a Higgs-mixed scalar decays to leptons are more important than reabsorption. For $m_\phi>2m_\mu$  decays to muons are fast, which leads to the right-hand edge of the constrained region being almost vertical.

For comparison, in Figure~\ref{fig:Higgs-portal results} we also show collider limits on Higgs-mixed scalars \cite{Egana-Ugrinovic:2019wzj,Dev:2019hho,Dev:2021qjj,Balaji:2022noj} taken from Refs.~\cite{Balaji:2022noj,Lanfranchi:2243034} and the future sensitivity of the experiments DUNE \cite{Berryman:2019dme} and SHiP \cite{Lanfranchi:2243034}. For $m_\phi\lesssim 100~{\rm MeV}$ the existing bounds come dominantly from searches for rare kaon decays at NA62 \cite{Ruggiero:2020phq},  
for $200~{\rm MeV} \lesssim m_\phi \lesssim 300~{\rm MeV}$ they come from $\phi$ decays to leptons in the LSND experiment \cite{Foroughi-Abari:2020gju}, and for larger $m_\phi$ from LHCb \cite{LHCb:2015nkv}.  
The sensitivity of such searches depends on whether the scalar decays only to the SM (the constraints are plotted assuming this is the case); if it dominantly decayed to hidden sector particles the limits from rare meson decays would be slightly stronger \cite{Dev:2021qjj} while the LSND and LHCb limits weaken dramatically. 
Meanwhile, for $m_\phi\lesssim 0.1~{\rm MeV}$ there are strong cooling constraints from horizontal branch stars \cite{Hardy:2017,Knapen:2017xzo} (see also e.g. \cite{Bottaro:2023gep,Yamamoto:2023zlu}). 
We also indicate the part of parameter space in which the scalar's coupling to the SM requires that $m_\phi$ is fine-tuned even with a UV cutoff of order of a TeV.

It is important to note that there are generally significant bounds on theories involving new light scalars from cosmology, which are typically especially strong for $m_\phi \lesssim {\rm MeV}$. For example, a single new scalar mixing with the Higgs is strongly constrained because too many late decays of the thermal population of the scalar to the SM disrupt the successful predictions of big bang nucleosynthesis \cite{Krnjaic:2015mbs,Ibe:2021fed,Fradette:2018hhl}. 
We do not plot these limits because these are very sensitive to the details of a particular theory.

Our constraints on $\sin\theta$ at $m_\phi \gtrsim 10~{\rm MeV}$, where continuum emission dominates, are about an order of magnitude weaker than those of Ref.~\cite{Balaji:2022noj}, which calculated the emission rate using the OPE approximation. For such $m_\phi$ the OPE approximation predicts that emission is dominantly from the external nucleons so this is a direct comparison, see Appendix~\ref{appendix: OPE comparison} for further details. As discussed in the introduction, in the case of axion emission the inclusion of higher corrections to the OPE calculation leads to results that are weaker than from the OPE approximation and which are similar to those obtained using the soft approximation. 
We therefore expect our constraints to be conservative without being overly cautious. At smaller $m_\phi$ Ref.~\cite{Balaji:2022noj} considered emission from the pion propagator still in the OPE approximation, obtaining limits that are about an order of magnitude stronger than ours from resonant emission.  
It is reasonable to expect that the OPE approximation might overestimate the production rate at small as well as large $m_\phi$, so we again regard our constraint as a useful robust limit. It might be interesting to analyse higher order chiral perturbation theory corrections to the OPE approximation for the production of a scalar, but we leave this as possible future work. 
We also note that if there  was substantial absorption onto the internal pion propagator the more efficient trapping could weaken our bounds at larger $\sin\theta$, although this is unlikely to be important for a Higgs-mixed scalar given the collider limits.

\subsection{Leptophilic and Nucleophilic Scalars}\label{ss:lepto}

\begin{figure}
    \centering
{{\includegraphics[width=0.49\textwidth]{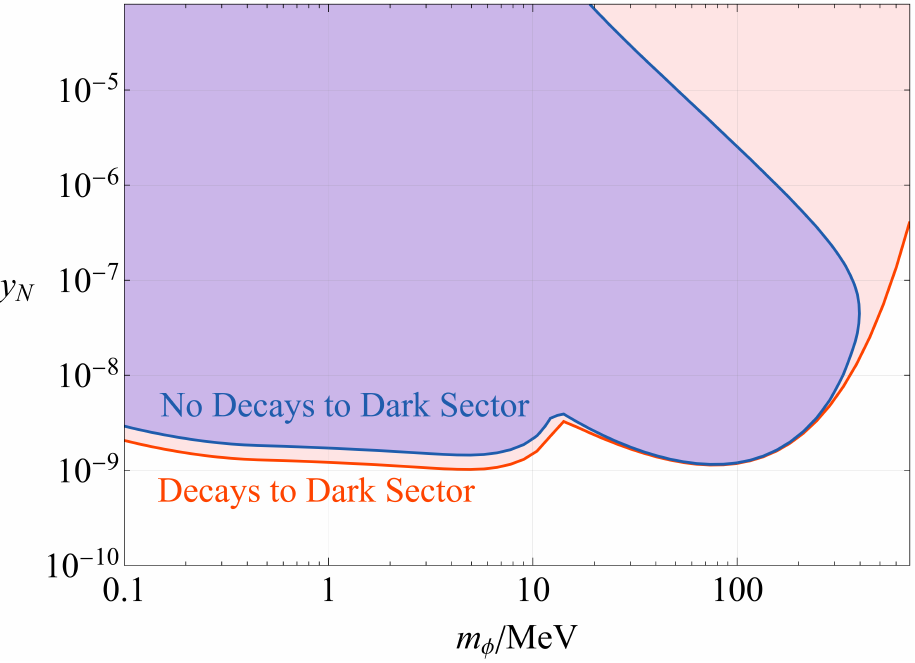} }}
{{\includegraphics[width=0.49\textwidth]{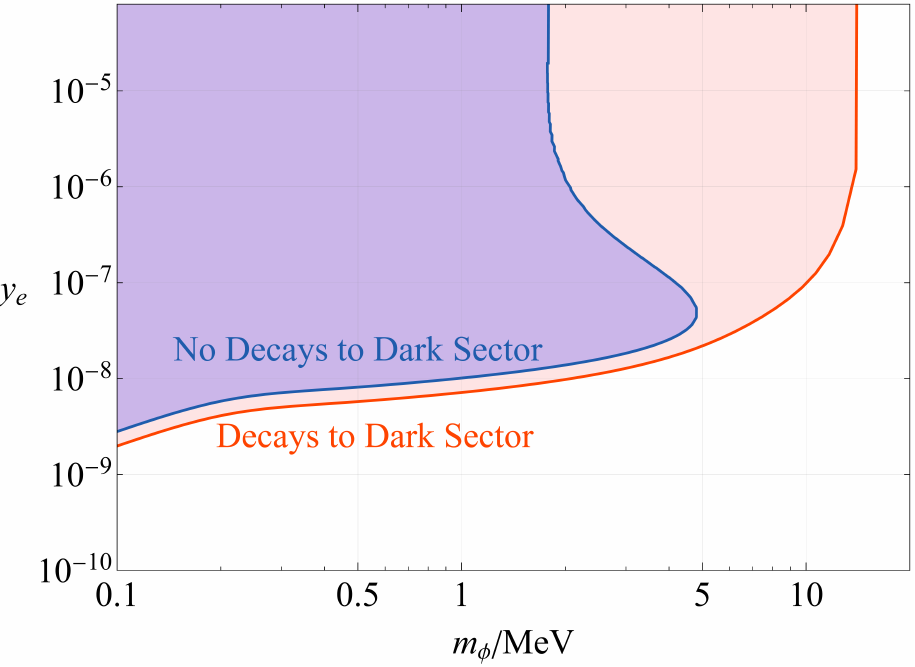} }}
    \caption{ \textit{\textbf{Left:}}
    Supernova constraints on a scalar $\phi$ that couples only to nucleons $N$ via an interaction $y_N \phi \bar{N}N$. Results are shown for theories in which $\phi$ decays quickly to some additional dark sector particles that do not interact with the Standard Model (``Decays to Dark Sector") and also assuming there are no such decays (``No Decays to Dark Sector"). 
    \textit{\textbf{Right:}} The analogous constraints on a scalar that couples only to leptons with coupling constant $y_e$, considering only resonant production.}
    \label{Fig: Lepto and Nucleo bounds}
\end{figure}

We can immediately adapt our calculations to the case of a nucleophilic scalar, i.e. one that couples only to nucleons $N$ through an interaction $y_N \phi \bar{N}N$. The production processes are identical to those for a Higgs-mixed scalar, but for all relevant $m_\phi$ there are no significant decays so the upper edge of the constrained region is determined only by reabsorption. The resulting limit is plotted in Figure~\ref{Fig: Lepto and Nucleo bounds} left. 

We also show the supernova constraint obtained in the absence of reabsorption, as would be applicable to a theory in which, after production, the scalar quickly decays to other hidden sector states that are decoupled from the SM (Figure~\ref{Fig: Lepto and Nucleo bounds} right). In particular, we consider the idealised scenario that the scalar decays instantaneously to some additional (unspecified) dark sector particles that have vanishing interaction with the standard model. In a realistic theory, one would need to take into account $\phi$'s finite lifetime as well as additional complications such as the new dark sector particles themselves having interactions with the standard model. 
In our limit of instantaneous decays, the constraints extend to such large $m_\phi$ that the soft approximation is questionable, so we cut the plot and stress that caution is required in this part of parameter space.  At weak coupling for $m_\phi\lesssim 20~{\rm MeV}$ there is a small difference between the constraints with and without fast decays to a hidden sector because the inward-going $\phi$ are resonantly reabsorbed in the former case despite $y_N$ being small. Collider constraints will again be relevant to these theories, however we do not plot them because they depend on the details on the underlying theory that gives rise to the interaction with nucleons \cite{Knapen:2017xzo}.

By considering mixing induced by electrons, we can also calculate the rate of resonant production of leptophilic scalars with an interaction $y_e\phi\bar{e}e$. We plot the resulting constraints in Figure~\ref{Fig: Lepto and Nucleo bounds} right.  In this case, the constraints on $y_e$ are sufficiently weak (i.e. the relevant values of $y_e$ are sufficiently large) that if $\phi$ decays only to SM particles then for $m_\phi\gtrsim 2\tilde{m}_e$ decays to electrons are fast enough that the Raffelt criterion leads to almost no bound. Had we used the vacuum electron mass the edge of the constrained region for $y_e\gtrsim 10^{-6}$ would be slightly shifted to $2m_e$. 
We again also show the limit obtained assuming the scalar decays quickly to hidden sector states, in which case energy can efficiently escape the protoneutron star even for $m_\phi\gg 2m_e$. Similar bounds on leptophilic scalars from resonant production were obtained in Ref.~\cite{Knapen:2017xzo}, which however used the thermal electron mass rather than the more appropriate vacuum mass to calculate the mixing (this was noted in \cite{DeRocco:2022rze}). In addition to resonant production, there will be other production processes relevant to such a leptophilic scalar, with semi-Compton scattering expected to be the most important. Inclusion of these could significantly strengthen the  supernova constraints, but we leave a dedicated analysis to future work.

\section{Discussion} \label{sec:discussion}

In this paper we have studied supernova constraints on new light scalars. Given our aim of reliable bounds, we have focused on production processes that are robust against uncertainties in the nucleon scattering cross section. The combination of resonant emission (relevant for fairly small $m_\phi < \omega_p \simeq 10~{\rm MeV}$) and soft emission (which makes use of experimental data to evade nuclear physics uncertainties and is relevant for larger $m_\phi$) enables limits to be obtained for all scalar masses for which supernova bounds are relevant $m_\phi\lesssim 200~{\rm MeV}$. These bounds are nicely complementary to the region of parameter space that will be probed by future experiments such as DUNE and SHiP, which are most sensitive for $m_\phi\gtrsim 200~{\rm MeV} $ \cite{Dev:2021qjj,Lanfranchi:2243034}. 
However, as discussed in detail in Section~\ref{sec:supernovae}, further observations of the remnant of SN~1987A and developments in simulations would be extremely useful for confirming the energy loss criterion. Ultimately, full supernova simulations incorporating the new scalar (analogous to those attempted in  Refs.~\cite{Keil:1996ju,Hanhart:2001fx,Fischer:2016cyd}) will be the best route to precise bounds, and our calculation of resonant emission, and refinement of other points such as degeneracy effects, would provide important inputs for these. Connected to this, it is plausible that for large $\sin\theta$ the additional energy transport by the new scalar within the neutrinosphere is not harmless and there could be constraints beyond the Raffelt bound relevant to the trapping regime (moreover, the supernova profiles we have used are probably not trustworthy for large $\sin\theta$ so our limits in this regime should anyway be interpreted with caution). 

There are various possible extensions to our work. 
For example, we did not take into account thermal and in-medium corrections to the nucleon dispersion relations or residue factor, which may be important. Specifically, it is known that in-medium effects could significantly affect nucleon masses reducing them by roughly $10\% - 40\%$ (see~\cite{Steiner:2012rk}, as well as a recent work incorporating this effect into the supernova cooling constraints for BSM particles~\cite{Carenza:2019pxu}). We expect that a careful analysis of this effect might strengthen the inferred constraints.
As mentioned, it would also be useful to calculate the rate of leptophilic scalar emission by semi-Compton processes. Given recent suggestions that the muon abundance in supernovae might be substantial \cite{Bollig:2017lki} it could also be interesting to investigate whether resonant production of a scalar that couples dominantly to muons might be possible (see Refs.~\cite{Croon:2020lrf,Caputo:2021rux,Alda:2024cxn} for related work).

Although we have focused on the impact of the produced scalars on the evolution of a supernova and in particular the resulting neutrino burst, in specific theories there could be other signals. For example, it might be possible to detect the products of decays of the scalar to SM particles outside the neutrinosphere (analogous decays of BSM particles to photons and neutrinos have previously been considered, see e.g. Refs.~\cite{Hannestad:2001jv,Kazanas:2014mca,Jaeckel:2017tud,DeRocco:2019njg,Fiorillo:2022cdq,Diamond:2023scc}).
Moreover, besides detecting the products of such decays directly, it might be possible to infer their presence by the energy they transfer to the supernova envelope. Indeed, the observed supernovae luminosities have been used to set limits on various feebly-interacting particles in the past~\cite{Sung:2019xie, Caputo:2022mah}. 
A detailed analysis of such signals could lead to additional constraints on new light scalars, or, if we are fortunate enough to observe a new nearby supernova, an opportunity to discover BSM physics. 

Finally, it could be interesting to investigate whether mixing with the longitudinal photon could affect the cosmological dynamics of new light scalars. For example, resonant conversion, which would occur at particular times in the early Universe, might have an impact on the scalar's abundance (similar effects have been studied for dark photon dark matter \cite{Redondo:2008ec}). 

\section*{Acknowledgements}

EH acknowledges the UK Science and Technology Facilities Council for support through the Quantum Sensors for the Hidden Sector collaboration under the grant ST/T006145/1 and UK Research and Innovation Future Leader Fellowship MR/V024566/1. AS is funded by the UK Research and Innovation grant MR/V024566/1. HS is supported by The Science and Technology Facilities Council under the grant ST/X508664/1.
We acknowledge the COST Action COSMIC WISPers CA21106, supported by COST (European Cooperation in Science and Technology).
For the purpose of open access, the author has applied a CC BY public copyright license to any Author Accepted Manuscript (AAM) version arising from this submission.

\appendix

\section{Theoretical setup}\label{appendix: Theoretical setup}

In this Appendix we summarise the theoretical foundations underlying our work by outlining the derivation of the production rate directly from the density matrix formalism, following \cite{Laine:2016}. Our setup consists of a bath of SM particles in thermal equilibrium and a weakly interacting scalar field $\phi$, such that initially, before any scalars are produced, the density matrix can be decomposed as a thermal distribution  for the SM particles and a pure vacuum state for the scalars
\begin{equation}\label{eqn: initial density matrix}
    \hat{\rho}(0) = \hat{\rho}_{\text{bath}}\otimes|0\rangle \langle 0 |~.
\end{equation}
The density matrix is then evolved in time by the Hamiltonian $\hat{H}$ according to the Liouville-von Neumann equation
\begin{equation}\label{eqn: Liouville von Neumann}
    i \frac{d\hat{\rho}(t)}{dt} = [\hat{H},\hat{\rho}(t)]~.
\end{equation}
This can then be used to find the phase space density of scalar particles:
\begin{equation}\label{pspden}
    \mathcal{F}_\phi(t,\boldsymbol{k}) = \frac{(2\pi)^3}{V} \text{Tr}\left[\hat{a}^\dag_{\boldsymbol{k}}\hat{a}^{}_{\boldsymbol{k}}\hat{\rho}(t)\right]~,
\end{equation}
where $\hat{a}^\dag_{\boldsymbol{k}}$ is the creation operator for a scalar of momentum $\boldsymbol{k}$, and $V$ is the volume of the considered region of the SM plasma. The full solution to this system of equations would describe the evolution of any particle being produced and reaching equilibrium with the existing heat bath, however we will restrict ourselves to times much shorter than the equilibration time, such that the density matrix never differs much from the initial state given in Eq.~\eqref{eqn: initial density matrix}. This approximation is justified by the assumed weakness of the interactions between the scalar and the SM particles: the scalar leaves the production region before it can come into equilibrium with it. We solve perturbatively in the coupling $g_f$ between the scalar and the SM and therefore split the Hamiltonian into parts $\hat{H} = \hat{H}_{\text{bath}}+\hat{H}_{\phi}+\hat{H}_{\text{int}} = \hat{H}_{0}+\hat{H}_{\text{int}}$, where $\hat{H}_{\text{int}}= \int d^3\boldsymbol{x} \ \hat{\phi}(x) \hat{J}(x)$ with the SM particle current $\hat{J}$ from Eq.~\eqref{smcurrent} for our case of interest. The thermal density matrix for the SM particles is given by $\hat{\rho}_{\text{bath}} = e^{-\beta \hat{H}_{\text{bath}}}/Z_{\text{bath}}$, where $Z_{\text{bath}}$ is the partition function of the SM bath. Substituting this into Eq.~\eqref{eqn: initial density matrix} and using it as an initial condition for Eq.~\eqref{eqn: Liouville von Neumann}, one can obtain a solution to the interaction-picture version of Eq.~\eqref{eqn: Liouville von Neumann} by means of a Dyson series. Since $\langle 0 |\hat{a}^\dag_{\boldsymbol{k}}\hat{a}^{}_{\boldsymbol{k}}| 0\rangle = 0$, and an odd total number of creation/annihilation operators under the trace gives 0 as well, the lowest order term in the Dyson series that yields a non-zero result when substituted into the phase space density Eq.~\eqref{pspden} is quadratic in $\hat{H}_{\text{int}}$, and thus also in $\hat{J}$. In particular, using $\langle 0 |\hat{a}^{}_{\boldsymbol{r}} \hat{a}^\dag_{\boldsymbol{k}}\hat{a}^{}_{\boldsymbol{k}} \hat{a}^\dag_{\boldsymbol{p}}| 0\rangle = \delta^{(3)} (\boldsymbol{r} - \boldsymbol{k})\, \delta^{(3)} (\boldsymbol{p} - \boldsymbol{k})$ and performing the integrals over the momenta coming from the Fourier decomposition of two $\hat{\phi}$ insertions, one finds that the rate of scalar particle production is given by:
\begin{equation}
    \frac{d}{dt}{\mathcal{F}}_\phi(t,\boldsymbol{k}) = \frac{1}{2V \omega} \int_0^t dt' \int d^3 \boldsymbol{x} \int d^3 \boldsymbol{y} \left\langle \hat{J} (y) \hat{J} (x) e^{iK\cdot (x-y)} + \hat{J} (x) \hat{J} (y) e^{iK\cdot (y-x)} \right\rangle_{\beta} \; + \; O(g_f^4) \, ,
\end{equation}
where $x = (t,\boldsymbol{x})$, $y = (t',\boldsymbol{y})$, $K = (\omega, \boldsymbol{k})$ is the 4-momentum of the produced scalar, $\omega = \sqrt{\boldsymbol{k}^2 + m_{\phi}^2}$ is its energy, and $\left\langle \dots \right\rangle_{\beta} \equiv \text{Tr} \left( \dots \hat{\rho}_{\text{bath}} \right)$ denotes thermal averaging over the SM bath. Sending $t \to \infty$ as suggested by the fact that we are interested in the emission on time scales much longer than the time scale of the interactions within the SM bath, and using $\langle \hat{J} (x) \hat{J} (y) \rangle_{\beta} = \int d^4 Q\, \exp (iQ\! \cdot \! (x-y))\, \pi^{<} (Q)$ as follows from the definition Eq.~\eqref{eq:pi_J}, one can easily perform the remaining space and time integrals, and obtain the final result for the production rate:
\begin{eqnarray}
    \dot{\mathcal{F}}_\phi( \boldsymbol{k} ) \; = \; \frac{1}{2\omega} \pi^{<} (\omega, \boldsymbol{k}) +  O(g_f^4) \, ,
\end{eqnarray}
which matches Eq.~\eqref{eq:GammaProd} of the main text.

In practice, the source $\hat{J}(x)$ typically contains many terms involving different coupling constants for each of the Standard Model field that $\phi$ interacts with. The notation $\mathcal{O}(g_{f}^4)$ in Eq.~\eqref{eq:GammaProd} denotes anything quartic or higher in any combination of these couplings.

\section{Proton and electron mediated mixing}\label{appendix: Relativistic deg mixing}
\begin{figure}
    \centering
    \begin{tikzpicture}
        \begin{feynman}[large]
            \vertex (a){$1$};
            \vertex [right=of a] (b);
            \vertex [above right=0.1em of b] {$1$};
            \vertex [below right=0.1em of b] {$1$};
            \vertex [right=of b] (c);
            \vertex [above left=0.1em of c] {$1$};
            \vertex [below left=0.1em of c] {$1$};
            \vertex [right=of c] (d) {$\mu,1$};

            \diagram* {
              (a) -- [scalar] (b),
              (b)--[fermion,half left,momentum=$K+P$](c)--[fermion,half left,momentum=$P$](b),
              (c)--[photon](d)
                };
        \end{feynman}
    \end{tikzpicture}
    \caption{The one-loop diagram contributing to the mixing self-energy. Lorentz indices are labeled as well as time contour indices. The internal fermion is either an electron or a proton in the cases of interest.}
    \label{fig:1 loop mixing self energy}
\end{figure}
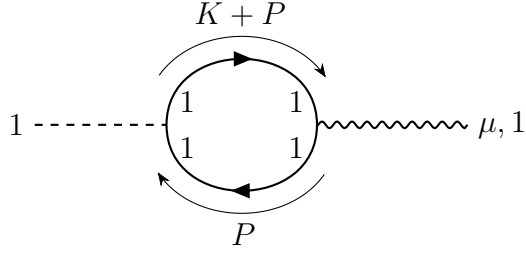

Here we present a real-time thermal field theory calculation of $\Pi_{\phi L}^{RA}$, considering the contributions from both (relativistic and degenerate) electrons and (non-relativistic) protons, typical of the supernova medium. We also provide some physical interpretation for the results. We begin by considering the real part; this is most easily obtained by changing to the 1/2 basis using the relation $\text{Re}[\Pi_{\phi L}^{RA}] = \text{Re}[\Pi_{\phi L}^{11}]$, which can be shown using the spectral representation of the self-energies \cite{Gelis:1997}. Since the only non-vanishing vertex index assignments are $g_{111}$ and $g_{222}$, there is only one diagram to consider, shown in Figure~\ref{fig:1 loop mixing self energy}. We then require the $11$-component of the fermion propagator, which is given by 
\begin{equation}
    D^F_{11}(Q) = (\slashed{Q}+m_f)(\Delta^F(Q)+\mathcal{S}(Q))~,
\end{equation}
where $m_f$ is the fermion mass, $Q = (q_0, \boldsymbol{q})$ is its 4-momentum, $\Delta^F(Q)$ is the usual Feynman propagator, and $\mathcal{S}(Q)$ is shorthand for the ``statistical" part
\begin{equation}
    \mathcal{S}(Q) = 2\pi \eta \delta(Q^2-m_f^2)(\theta(q_0)n(x_q)+\theta(-q_0)n(-x_q))~,
\end{equation}
which has the effect of putting the propagator on-shell and is accompanied by a thermal distribution function. Here we have introduced the compact notation $\eta = +1(-1)$ for bosons (fermions) and 
\begin{equation}
    n(x_q)=\frac{1}{e^{x_q}-\eta}~, \qquad x_q = \beta(q_0-\mu_{f})~,
\end{equation}
where $\beta$ is the inverse temperature and $\mu_f$ is the fermion chemical potential.
This allows us to write down an expression for the mixing self-energy
\begin{equation}\label{eqn:mixing self energy setting up integral}
\begin{split}
     -i \Pi^{11}_{\phi\gamma, \mu}(K) = -\int\frac{d^4P}{(2\pi)^4}&(-ie)(-ig_{f})\Tr\left[(\slashed{K}+\slashed{P}+m_f)(\slashed{P}+m_f)\gamma_\mu\right] \\ \times&\left(\Delta^F(K+P)+\mathcal{S}(K+P)\right)\left(\Delta^F(P)+\mathcal{S}(P)\right)~,
\end{split}
\end{equation}
where $K = (k_0, \boldsymbol{k})$ is the external 4-momentum and $P = (p_0, \boldsymbol{p})$ is the 4-momentum running in the loop.
There are three types of contribution to this expression: those of the form $\Delta^F\Delta^F$, which correspond to the mixing in vacuum and must therefore vanish by Furry's theorem, those of the form $\mathcal{S}\mathcal{S}$, which only contribute to the imaginary part and so we discard them, and cross terms of the form $\Delta^F \mathcal{S}$. By using the delta functions to simplify the phase space integration and using the relation $-x_{-k} = -\beta(-k_0-\mu_f) = \beta(k_0+\mu_f)$ to introduce the statistical distributions for fermions $\kappa_f(E) \equiv n(x_E)$ and antifermions $\kappa_{\bar{f}}(E) \equiv n(-x_{-E})$, we arrive at
\begin{equation}\label{eqn: Mixing self energy}
        \text{Re } \Pi_{\phi \gamma,\mu}^{11}(K) =4 g_f e m_f\text{ Re }\int\frac{d^3\boldsymbol{p}}{(2\pi)^3}\frac{1}{2E}\left(\kappa_f(E)-\kappa_{\bar{f}}(E)\right)\frac{(P\cdot K)K_\mu-K^2 P_\mu}{(P\cdot K)^2-(K^2)^2/4}~,
\end{equation}
where the 4-momentum $P = (E,\boldsymbol{p})$ is now on-shell so that $E = \sqrt{\boldsymbol{p}^2+m_f^2}$. This is consistent with the results found in the imaginary-time formalism \cite{DeRocco:2022rze}, as can be inferred from the general relation between the real parts of self-energy in imaginary and real-time formalisms: $\text{Re}\, \Pi_{11} (K) = \text{Re}\, \Pi (k_0 + i\epsilon k_0, \mathbf{k})$, where $\Pi$ is the analytically continued imaginary-time self-energy.

The mixing self-energy has a number of interesting features. Firstly we can easily see that it obeys the Ward identity $K^\mu \Pi^{\phi\gamma}_\mu(K)=0$. It is also straightforward to show that it vanishes on contraction with a transverse polarization tensor $\Pi^{\phi\gamma}_\mu(K) \epsilon_{\text{T}}^\mu(K) = 0$. This is unsurprising since it is only the longitudinal mode of the photon that has the right quantum numbers to mix with the scalar. A final interesting feature is that the mixing vanishes in the case that the medium has equal numbers of particles and antiparticles. This can be understood in several ways, for example by considering charge conjugation symmetry. The photon is odd under charge conjugation, whereas the scalar is even, and therefore for the two to mix the medium must explicitly break charge conjugation symmetry by having an uneven number of particles and antiparticles (an alternative justification was provided in Ref. \cite{DeRocco:2022rze}). Physically, the mixing can be interpreted as forwards semi-Compton scattering off of the particles in the medium, and Eq.~\eqref{eqn: Mixing self energy} can be found by a thermal average over this matrix element. This is exactly analogous to how the photon self-energy can be interpreted as forward scattering on the constituents of the medium \cite{Raffelt:1996wa}. 

The two charged particles present in the supernova core that may run in the thermal loop are protons and electrons. The protons are non-relativistic (but not necessarily non-degenerate), and the integral of Eq.~\eqref{eqn: Mixing self energy} can be performed in this regime~\cite{Hardy:2017} to give 
\begin{equation}
    \text{Re}[\Pi_{\phi L}^{RA}(K)]_{\text{protons}} = g_P e\frac{k \sqrt{K^2}}{\omega^2}\frac{n_P}{m_P}~,
\end{equation}
where $k = |\boldsymbol{k}|$ and we have defined the scalar-longitudinal photon mixing as $\Pi_{\phi L}^{RA} = \epsilon^{\mu}_L\Pi_{\phi \gamma,\mu}^{RA}$, with $\epsilon^{\mu}_L$ a longitudinal polarization vector. On the other hand, the electrons are degenerate and ultra-relativistic, so performing the integral in this regime yields \cite{DeRocco:2022rze}
\begin{equation}\label{eqn: electron mixing unsimplified}
    \text{Re}[\Pi_{\phi L}^{RA}(K)]_\text{electrons} = \frac{g_e e}{\pi^2}\frac{\sqrt{K^2}}{k}m_e \mu_e\left(\frac{k_0}{2 k}\log\left(\frac{k_0+k}{k_0-k}\right)-1\right)~.
\end{equation}
In both cases the $(K^2)^2/4$ term is dropped from the denominator of Eq.~\eqref{eqn: Mixing self energy} before the integral is performed. Following a similar logic to that presented in Ref.~\cite{Braaten:1993jw}, we can argue that $(P\cdot K)^2\gg K^4/4$ by noticing that the expression will be evaluated on the scalar's mass-shell so $K^4 = m_\phi^4$, and that the phase space integral will be dominated by a particular value of $P$, corresponding to non-relativistic protons with $P \approx (m_P,0)$ or relativistic degenerate electrons at the Fermi energy with $P\approx(E_F, \boldsymbol{p}_F)$. It is then straightforward to check that the $K^4$ term is negligible in our cases of interest. This has the effect of making the integral in Eq.~\eqref{eqn: Mixing self energy} real.

When evaluating Eq.~\eqref{eqn: electron mixing unsimplified} on-resonance, we can make use of the photon on-shell condition $K^2 = \text{Re}\,\Pi_L(K)$ with the real part of self-energy given by Eq.~\eqref{eqn: photon self energy}, as well as the scalar on-shell condition, $K^2 = m_\phi^2$, and the expression for the plasma frequency as given by Eq.~\eqref{eqn: plasma frequency} (in the limit where the electron chemical potential is much greater than the temperature), to simplify Eq.~\eqref{eqn: electron mixing unsimplified} to 
\begin{equation}
    \text{Re}[\Pi_{\phi L}^{RA}(\omega_*)]_\text{electrons} = \frac{g_e m_\phi m_e k_*}{e\mu_e}~,
\end{equation}
where $\omega_*$ denotes the resonant frequency and $k_*$ the corresponding momentum.

We can easily compare the relative sizes of the mixing due to protons and electrons in the case of a Higgs-mixed scalar:
\begin{equation}
    \left(\frac{\Pi_\text{electrons}}{\Pi_\text{protons}}\right)^2 = \frac{g_e^2}{g_P^2 e^4}\left(\frac{m_e m_P}{n_P \mu_e}\right)^2\omega_*^4\sim 10^{-6}\left(\frac{\omega_*}{T}\right)^4~,
\end{equation}
where in the final approximation we used the values of temperature, proton density, and electron chemical potential from the $20~M_\odot$ progenitor at a radius of $10~{\rm km}$ from the centre. The protons dominate the mixing due to their larger coupling with the Higgs-mixed scalar and their large number density in the core of the supernova. Since most resonant production will indeed occur near the centre of the protoneutron star, the protons will dominate the mixing for all cases of interest (it is possible that electrons might be relevant for very low mass scalars, $m_\phi\sim 0.5~{\rm MeV}$, in the trapping regime when resonance from the outer regions is relevant, but such parameter space is already excluded considering only the mixing from protons).

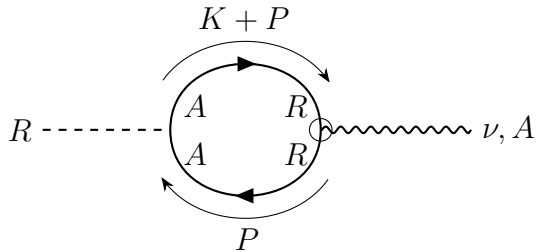
\begin{figure}
\centering
    \begin{tikzpicture}
        \begin{feynman}[large]
            \vertex (a){$R$};
            \vertex [right=of a] (b);
            \vertex [above right=0.1em of b] {$A$};
            \vertex [below right=0.1em of b] {$A$};
            \vertex [right=of b] (c);
            \vertex [above left=0.1em of c] {$R$};
            \vertex [below left=0.1em of c] {$R$};
            \vertex [right=of c] (d) {$\nu,A$};

            \draw (c) circle[radius=0.15cm];
            
            \diagram* {
              (a) -- [scalar] (b),
              (b)--[fermion,half left,momentum=$K+P$](c)--[fermion,half left,momentum=$P$](b),
              (c)--[photon](d)

                };
        \end{feynman}
    \end{tikzpicture}
    \caption{The cut one-loop diagram contributing to the mixing self-energy. Lorentz indices are labeled as well as time contour indices. The internal fermion is either an electron or a proton in the cases of interest.}
    \label{fig: 1 loop RA for cutting}
\end{figure}

Having discussed the real part of the mixing self-energy in detail, we now turn to the imaginary part, with the aim of justifying our assumption that $\text{Im}[\Pi_{\phi L}^{RA}]\ll\text{Re}[\Pi_{\phi L}^{RA}]$. We will do this by applying the cutting rules of Bedaque, Das and Naik~\cite{Bedaque:1996af} in the $R/A$ basis as described in Ref.~\cite{Gelis:1997} to the 1-loop self-energy diagram. The standard prescription following Kobes and Semenoff~\cite{Kobes:1985kc} is that vertices may be either ``circled" or ``uncircled", which come with different vertex factors, and the propagators change depending on what types of vertices they connect. We must then sum over all ways of circling vertices except the cases where all or none of the vertices are circled. While this sounds like a terrible increase in complexity, in fact in the $R/A$ basis and for the 1-loop self-energy diagram we are left with only a single non-vanishing diagram to evaluate, shown in Figure~\ref{fig: 1 loop RA for cutting}. We then need the ``cut" propagators
\begin{equation}
    D_{A\underline{R}}(Q) = -D_{\underline{R}A}(Q)  = (\slashed{Q}+m_f)2\pi\varepsilon(q_0)\delta(Q^2-m_f^2)~,
\end{equation}
where underline means that the corresponding vertex is circled, and $\varepsilon(q_0) \equiv \text{sgn} (q_0)$. We also need the vertex factors \cite{Gelis:1997,van_Eijck:1994}
\begin{equation}
    g_{\underline{A}\underline{R}\underline{R}} = ig_{f}, \quad g_{RAA}(p_1,p_2,p_3) = ig_{f}(1+\eta_2n(x_{p_2})+\eta_3n(x_{p_3}))~. 
\end{equation}
Here, $x_{p_i}$ are determined by treating the chemical potential analogously to momentum, that is ensuring it is conserved at every vertex. All momenta should be taken to be flowing in to the vertex. The imaginary part of the diagram is then given by $-1/2$ of the sum over all the cut diagrams. In our case this gives
\begin{equation}\label{eq46}
\begin{split}
    \text{Im} \left[ \Pi_{\phi \gamma,\mu}^{RA}(K) \right] = \frac{1}{2}(ig_{f})(ie)\int\frac{d^4P}{(2\pi)^4}&\Tr\left[(\slashed{K}+\slashed{P}+m_f)(\slashed{P}+m_f)\gamma_\mu\right]\\
    \times&(1-n(\beta(p_0-\mu_{f}))-n(\beta(-k_0-p_0+\mu_{f})))\\
    \times&2\pi\varepsilon(k_0+p_0)\delta((K+P)^2-m_f^2)\\
    \times&(-2\pi)\varepsilon(p_0)\delta(P^2-m_f^2)~.
\end{split}
\end{equation}
Eq.~\eqref{eq46} could be calculated in full generality. However, we are only interested in the behaviour of this near resonance, in which case $K$ must satisfy both the scalar and photon dispersion relations. Since the delta functions place both of the internal fermions on-shell, this now has identical kinematics to longitudinal photon decay or other processes like $f \gamma \rightarrow f$ which trivially have no phase space, and so we can use this to determine the phase space we must integrate over. For a loop of protons, the decay corresponds to $\gamma\rightarrow \bar{p}p$, which clearly has no accessible phase space for the temperatures and densities we consider. On the other hand a loop of electrons corresponds to $\gamma\rightarrow e^+e^-$. Naively this appears to be allowed, however if we were to use full thermal resummed electron propagators in the loop we would find that the electron picks up a thermal mass such that this decay is forbidden~\cite{Braaten:1990de}. Indeed, the in-medium photon dispersion relation $\omega (k)$ is bounded from above by the dispersion curve $\omega^2 = k^2 +3\omega^2_{p}/2$, while the in-medium electron dispersion relation $E(p)$ is bounded from below by the dispersion curve $E^2 = p^2 + m_{\text{eff}}^2$, where $\omega_{p}$ is the plasma frequency given in Eq.~\eqref{eqn: plasma frequency} and $m_{\text{eff}}$ is the in-medium contribution to the electron mass given subsequently in Eq.~\eqref{eq:me2}. If the decay is not possible for such bounding dispersion curves, i.e. if $2 m_{\text{eff}} > \sqrt{3} \omega_p / \sqrt{2}$, then it is also not possible for the physical case. In the region important for the resonant production of scalars, that is inside the core of the protoneutron star, $\mu_e \gg T$ and so $m_{\text{eff}} = \sqrt{3}\omega_p / 2$, which means that the decay of electromagnetic excitations into electrons is forbidden. Note that photon decay into the other in-medium electron modes known as ``plasminos" is exponentially suppressed at momenta higher than $m_{\text{eff}}$ due to the thermal residue factor~\cite{1992ApJ...392...70B}, while at momenta $p \lesssim m_{\text{eff}}$, an analysis of the dispersion relations analogous to the one above shows that photon decay into plasminos has no accessible phase space. We therefore see that there are no kinematically allowed contributions to the 1-loop mixing self-energy in the $R/A$ basis, and so $\text{Im}[\Pi_{\phi L}^{RA}]\ll\text{Re}[\Pi_{\phi L}^{RA}]$.

\section{In-medium decay rate}\label{appendix: In medium decay}

In this appendix we discuss the details of  scalar decays in the supernova environment and argue that the approximation made in the main text of using the vacuum rate in the case of a Higgs-mixed scalar is accurate enough for our purposes. 

From the perspective of thermal field theory, the scalar's in-medium decay rate is related to the imaginary part of the first diagram in Figure~\ref{fig:Lowest order loops}.  
Indeed, if this is calculated using the cutting rules presented in \cite{Gelis:1997} one recovers the result obtained by using the vacuum matrix element and accounting for Pauli-blocking of the final state electrons and positrons. However, to get the true decay rate we would have to use the full resummed electron propagator in the loop diagram. Much like the case of the photon discussed in Section~\ref{sec: resonant production}, the electron gets a thermal self-energy. One physical effect of this is to give the electron a complicated dispersion relation~\cite{Le_Bellac:1996,Kapusta:2006pm} with energy and momentum expressed only parametrically. Since this dispersion relation is not Lorentz invariant, integrating over the kinematically allowed phase space  would be very computationally challenging. 
The situation is worsened by the emergence of collective fermionic modes known as plasminos. These could provide another decay channel for the scalars with no vacuum analogue. On top of this, the electrons and plasminos would both pick up a residue factor due to their shifted pole compared to the vacuum propagator (this is sometimes referred to as a renormalization constant in analogy with wavefunction renormalization), which would alter the electron's interaction strength from its vacuum value.

\begin{figure}
    \centering
    \includegraphics[width=0.7\linewidth]{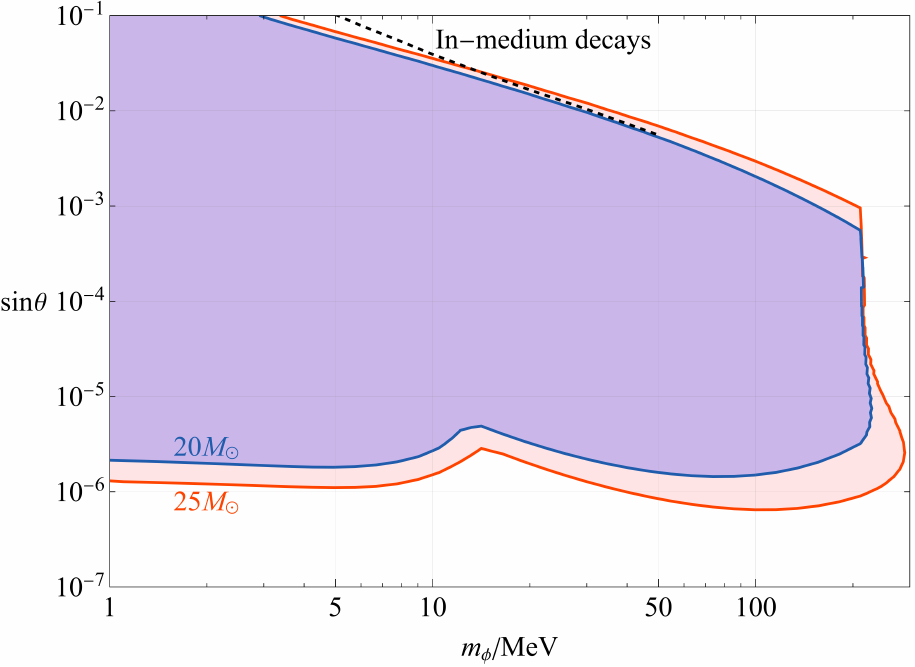}
    \caption{The impact of using the corrected in-medium electron decay rate (dashed black) on the constraints on a Higgs-mixed scalar $\phi$, compared to our main results obtained using the vacuum decay rate (solid blue). The corrected results are calculated for the $20~M_\odot$ progenitor. For comparison with the uncertainties arising from the choice of progenitor, we also show our main bounds for the $25~M_\odot$ progenitor (red).}
    \label{fig:comparing effects of in medium decays}
\end{figure}

To obtain a simple estimate of the impact that thermal corrections to the electron decay rate might have on our supernova constraints, we approximate the electron dispersion relation as $E^2 = p^2 + \tilde{m}_e^2$, with a thermal effective mass $\tilde{m}_e^2$ given by \cite{1992ApJ...392...70B}
\begin{equation} \label{eq:me1}
    \tilde{m}_e = \frac{m_e}{2}+\sqrt{\frac{m_e^2}{4}+m_\text{eff}^2}~,
\end{equation}
where
\begin{equation} \label{eq:me2}
    m_\text{eff}^2 = \frac{\pi \alpha}{2}\left(T^2 + \frac{\mu_e^2}{\pi^2}\right)~.
\end{equation}
This approximation matches the true dispersion relation in the limit where the electron momentum is much larger than the temperature and electron chemical potential (note that we use the thermal mass only in the dispersion relation; the vacuum electron mass is still used when calculating the spinor sums). 
Additionally, we take into account Pauli-blocking of the final state electrons and positrons, which we neglected in our main results. 
In Figure~\ref{fig:comparing effects of in medium decays} we show how these changes affect the SN~1987A bounds on a Higgs-mixed scalar. 
For large $m_\phi$ the electrons produced by the decays have energies much larger than $\tilde{m}_e$ and the Fermi energy and there is no change to the constraints. 
At small $m_\phi$, the bounds in the trapping regime are slightly strengthened compared to those obtained using the vacuum decay rate, because both Pauli-blocking and the effective electron mass reduce the decay rate allowing more scalars to escape the protoneutron star. The impact of the thermal effects is smaller than might be expected because decays are the dominant source of attenuation only in the outer regions of the protoneutron star where the electron thermal mass is fairly small. Given that the difference between constraints obtained by the two approaches only exceeds the uncertainties associated with the different progenitors for $\sin\theta\gtrsim0.03$ which is firmly excluded by collider limits, we use the much simpler vacuum decay rate for our main Higgs-mixed scalar results. As discussed in Section~\ref{ss:lepto}, for the case of a leptophilic scalar we use the effective mass, which slightly shifts the constraint.

\section{Comparison of the one-pion-exchange and soft approximations}\label{appendix: OPE comparison}
It has been argued that the one-pion-exchange calculation overestimates the bremsstrahlung production rate for neutrinos \cite{Hanhart:2000ae}, axions \cite{Carenza:2019pxu}, and dark photons \cite{Rrapaj:2015wgs} by up to an order of magnitude. Here we show that it also predicts a much higher bremsstrahlung production rate for scalars than our soft radiation approximation (SRA). We also discuss in detail the approximations and uncertainties associated with the SRA. 

In the OPE approximation, the scalar can be emitted either from the external nucleons or from the internal pion meaning that the matrix element squared is given by the Feynman diagrams
\begin{equation}\label{eqn: OPE diagrams}
|M|^2 = 
\left|\;
    \begin{tikzpicture}[baseline = {(0,0)},scale = 0.5]
        \tikzset{
            fermion/.style={draw=black, thick, postaction={decorate}, decoration={markings, mark=at position 0.5 with {\arrow{Straight Barb}}}}
        }

        \coordinate (n1) at (-2,1);
        \coordinate (n2) at (-2,-1);
        \coordinate (n3) at (2,1);
        \coordinate (n4) at (2,-1);
        \coordinate (v1) at (0,1);
        \coordinate (v2) at (0,-1);

        \coordinate (s1) at (1,1);
        \coordinate (s2) at (2,2);

        \draw[fermion] (n1) -- (v1) node[midway,above] {$N\;\;$};
        \draw[fermion] (v1) -- (n3);
        \draw[fermion] (n2) -- (v2) node[midway,above] {$N\;\;$};
        \draw[fermion] (v2) -- (n4);
        \draw[thick,dashed] (v1) -- (v2) node[midway,right] {$\pi$};

        \draw[dashed] (s1) -- (s2) node[left] {$\phi$};
        
    \end{tikzpicture}
    +
    \;
    \begin{tikzpicture}[baseline = {(0,0)},scale = 0.5]
        \tikzset{
            fermion/.style={draw=black, thick, postaction={decorate}, decoration={markings, mark=at position 0.5 with {\arrow{Straight Barb}}}}
        }

        \coordinate (n1) at (-2,1);
        \coordinate (n2) at (-2,-1);
        \coordinate (n3) at (2,1);
        \coordinate (n4) at (2,-1);
        \coordinate (v1) at (0,1);
        \coordinate (v2) at (0,-1);

        \coordinate (s1) at (1,-1);
        \coordinate (s2) at (2,-2);

        \draw[fermion] (n1) -- (v1) node[midway,above] {$N\;\;$};
        \draw[fermion] (v1) -- (n3);
        \draw[fermion] (n2) -- (v2) node[midway,above] {$N\;\;$};
        \draw[fermion] (v2) -- (n4);
        \draw[thick,dashed] (v1) -- (v2) node[midway,right] {$\pi$};

        \draw[dashed] (s1) -- (s2) node[left] {$\phi$};
        
    \end{tikzpicture}
    +
    \;
    \begin{tikzpicture}[baseline = {(0,0)},scale = 0.5]
        \tikzset{
            fermion/.style={draw=black, thick, postaction={decorate}, decoration={markings, mark=at position 0.5 with {\arrow{Straight Barb}}}}
        }

        \coordinate (n1) at (-2,1);
        \coordinate (n2) at (-2,-1);
        \coordinate (n3) at (2,1);
        \coordinate (n4) at (2,-1);
        \coordinate (v1) at (0,1);
        \coordinate (v2) at (0,-1);

        \coordinate (s1) at (-1,1);
        \coordinate (s2) at (0,2);

        \draw[fermion] (n1) -- (v1) node[midway,above] {$N\;\;$};
        \draw[fermion] (v1) -- (n3);
        \draw[fermion] (n2) -- (v2) node[midway,above] {$N\;\;$};
        \draw[fermion] (v2) -- (n4);
        \draw[thick,dashed] (v1) -- (v2) node[midway,right] {$\pi$};

        \draw[dashed] (s1) -- (s2) node[right] {$\phi$};
        
    \end{tikzpicture}
    +
    \;
    \begin{tikzpicture}[baseline = {(0,0)},scale = 0.5]
        \tikzset{
            fermion/.style={draw=black, thick, postaction={decorate}, decoration={markings, mark=at position 0.5 with {\arrow{Straight Barb}}}}
        }

        \coordinate (n1) at (-2,1);
        \coordinate (n2) at (-2,-1);
        \coordinate (n3) at (2,1);
        \coordinate (n4) at (2,-1);
        \coordinate (v1) at (0,1);
        \coordinate (v2) at (0,-1);

        \coordinate (s1) at (-1,-1);
        \coordinate (s2) at (0,-2);

        \draw[fermion] (n1) -- (v1) node[midway,above] {$N\;\;$};
        \draw[fermion] (v1) -- (n3);
        \draw[fermion] (n2) -- (v2) node[midway,above] {$N\;\;$};
        \draw[fermion] (v2) -- (n4);
        \draw[thick,dashed] (v1) -- (v2) node[midway,right] {$\pi$};

        \draw[dashed] (s1) -- (s2) node[right] {$\phi$};
        
    \end{tikzpicture}
    +
    \;
    \begin{tikzpicture}[baseline = {(0,0))},scale = 0.5]
        \tikzset{
            fermion/.style={draw=black, thick, postaction={decorate}, decoration={markings, mark=at position 0.5 with {\arrow{Straight Barb}}}}
        }

        \coordinate (n1) at (-2,1);
        \coordinate (n2) at (-2,-1);
        \coordinate (n3) at (2,1);
        \coordinate (n4) at (2,-1);
        \coordinate (v1) at (0,1);
        \coordinate (v2) at (0,-1);

        \coordinate (s1) at (0,0);
        \coordinate (s2) at (2,0);

        \draw[fermion] (n1) -- (v1) node[midway,above] {$N\;\;$};
        \draw[fermion] (v1) -- (n3);
        \draw[fermion] (n2) -- (v2) node[midway,above] {$N\;\;$};
        \draw[fermion] (v2) -- (n4);
        \draw[thick,dashed] (v1) -- (v2) node[midway,left] {$\pi$};

        \draw[dashed] (s1) -- (s2) node[right] {$\phi$};
        
    \end{tikzpicture}
    + \dots
    \right|^2,
\end{equation}
where $+\dots$ represents diagrams in the u-channel. Following Ref.~\cite{Dev:2020eam}, we split the energy production into that arising entirely from the first four terms, which we refer to as the ``external" contribution, that arising entirely from the final term, which we refer to as the ``pion" contribution, and that from cross terms which we refer to as the ``interference" contribution.

Meanwhile, in the SRA we assume non-relativistic nucleons and expand in the small parameters $a = m_\phi^2/(2m_N \omega)$ and $b = \slashed{K}/(2m_N)$, where $K = (\omega,\boldsymbol{k})$ is the 4-momentum of the produced scalar. While the SRA has the advantage of being independent from any nuclear physics uncertainties, there is a subtlety for scalars that does not arise in the dark photon or axion case. As first pointed out in Ref.~\cite{Dev:2020eam}, in the OPE approximation cancellations occur within the ``external" contribution to $\mathcal{O}(a^0)$. These cancellations also occur in the SRA meaning that the production rate vanishes at $\mathcal{O}(a^0b^0)$. We therefore work to $\mathcal{O}(a^1b^0)$, while throwing away terms $\mathcal{O}(a^0b^1)$, which are not calculable in the SRA. 
In the mass range where our bounds are set by the bremsstrahlung rate, $m_\phi\gtrsim 20$MeV, we expect $\omega\sim m_\phi$, and therefore $a\sim b$. It would therefore be surprising if the discarded $\mathcal{O}(a^0b^1)$ term led to order-of-magnitude corrections to our result. Furthermore, to leading order in chiral perturbation theory (i.e. in the OPE approximation) the cancellations extend to $\mathcal{O}(a^0b^1)$, providing some indication that this term may in fact be sub-leading to the $\mathcal{O}(a^1b^0)$ term that we have retained. 

\begin{figure}
    \centering
    \includegraphics[width=0.7\linewidth]{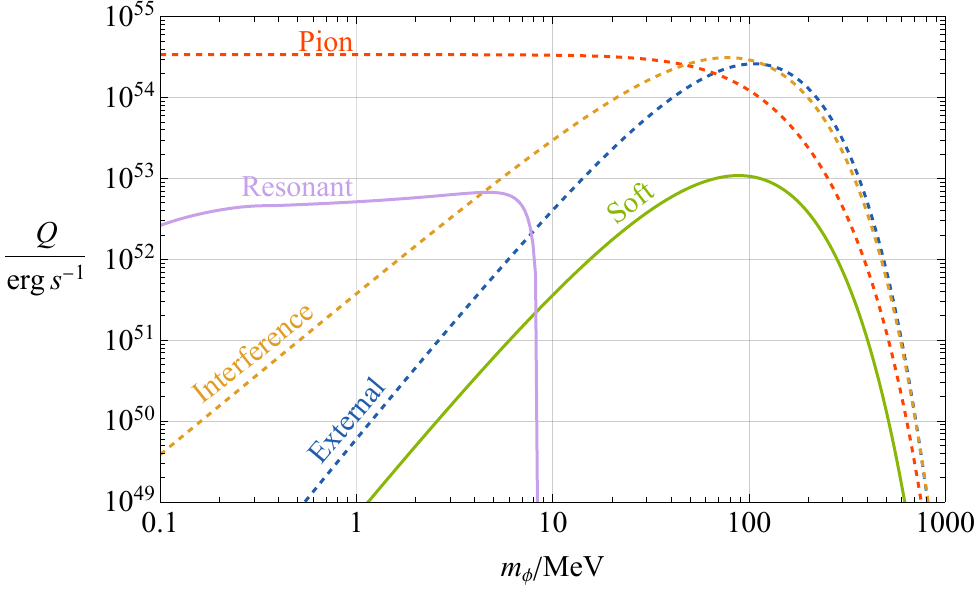}
    \caption{A comparison of the energy emission to a Higgs-mixed scalar with mass $m_\phi$ in the one-pion-exchange approximation (dashed) and by the processes that we use for our main results (solid). ``External" corresponds to scalar emission from external legs, ``Pion" corresponds to scalar emission from the internal pion propagator, and ``Interference" corresponds to the contribution from the interference of these two terms. ``Resonant" and ``Soft" correspond to the resonant production and soft bremsstrahlung production discussed in Section~\ref{sec:production}. For this comparison we fix $\sin\theta = 3\times 10^{-6}$ and use a constant temperature and density supernova profile.}
    \label{fig:OPE comparison}
\end{figure}

In Figure~\ref{fig:OPE comparison} we plot the energy emission into a Higgs-mixed scalar from a protoneutron star as a function of $m_\phi$. We show the three contributions from the OPE approximation separately, as well as the contribution from resonant production and the contribution from the SRA calculation of the bremsstrahlung rate. For this simple comparison we approximate the protoneutron star as  a uniform sphere of radius $10~{\rm km}$, temperature $41~{\rm MeV}$, baryon number density $6\times 10^5~ {\rm MeV}^3$, proton fraction $0.25$, and electron chemical potential $132~{\rm MeV}$ (corresponding to the values for the $20~M_\odot$ progenitor at a radius of $10~{\rm km}$). We also ignore attenuation due to reabsorption and decays, and the effects of degeneracy.

The soft approximation only captures emission from the external legs, and from Figure~\ref{fig:OPE comparison} we see that it predicts an emission rate around an order of magnitude lower than the corresponding contribution from the OPE calculation.  This is similar to previous results for axions, dark photons, and neutrinos. As a result, our main constraints for $m_\phi\gtrsim 20~{\rm MeV}$ are somewhat weaker than those obtained assuming the OPE approximation.

In the other mass range, $m_\phi\lesssim 20$MeV, the $m_\phi^2/(2m_N \omega)$ suppression is more substantial and we expect that emission from internal propagators could be more important than from the external legs. For example,  in the OPE approximation the ``pion" contribution becomes dominant, as shown in Figure~\ref{fig:OPE comparison}. 
Such emission from internal processes is not accounted for in the SRA, but fortunately for this mass range resonant emission gives a substantial contribution. Since the OPE approximation overestimates the ``external" contribution (which is captured by the SRA) by around an order of magnitude it seems unlikely that the true internal contribution is as large as the ``pion" contribution predicted in OPE. Indeed, the difference between the energy emitted from external production using the OPE and the soft approximation happens to be similar to that between the OPE internal production and resonant emission. We therefore view our results obtained from resonant production as a useful conservative limit. 
As mentioned in Section~\ref{sec:results}, in the future it might be interesting to carry out a higher order chiral perturbation theory calculation to further understand the significance of internal emission.

\section{Degeneracy effects on bremsstrahlung}\label{appendix: degeneracy bremsstrahlung}

In this Appendix we briefly outline the steps required to calculate the bremsstrahlung rate taking into account degeneracy and we analyse the impact of degeneracy on supernova constraints.

After substituting Eq.~\eqref{eqn: M in terms of cross section} into Eq.~\eqref{eqn: brem rate pre integratoin} we need to perform the phase space integration. First, we make the non-relativistic nucleon approximation $E_i \approx m_N$ and $E_\text{CM} \approx 2m_N$, and the ``no nucleon recoil" approximation which allows us to drop the scalar's 4-momentum from the delta function. Next, we change to centre of mass (CM) frame variables
\begin{equation}\label{eqn: brem change of vars}
    \boldsymbol{P} = \boldsymbol{p}_1 + \boldsymbol{p}_2, \quad
    \boldsymbol{P}' = \boldsymbol{p}_3 + \boldsymbol{p}_4, \quad
    \boldsymbol{q} = \frac{1}{2}(\boldsymbol{p}_1 - \boldsymbol{p}_2), \quad
    \boldsymbol{q}' = \frac{1}{2}(\boldsymbol{p}_3 - \boldsymbol{p}_4)~.
\end{equation}
The integral over $\boldsymbol{P}'$ can be performed immediately using the 3-momentum-conserving delta function. In the non-relativistic approximation, the energy-conserving delta function then reduces to 
\begin{equation}
    \delta\left(\frac{q^2}{m_N}-\frac{q'^2}{m_N}\right) = \frac{m_N}{2q}\delta(q-q')~,
\end{equation}
allowing the integral over $q' = |\boldsymbol{q}'|$ to be carried out. We note that the angular part of the integration over $\boldsymbol{q}'$ that remains is exactly the integral over the solid angle in the CM-frame, $d\Omega_{\text{CM}}$. We are left with an expression where the only dependence on the angle of $\boldsymbol{q}$ is contained within the initial Fermi factors
\begin{equation}
    f \left[\frac{1}{2}(\boldsymbol{P}+2\boldsymbol{q})\right] f \left[\frac{1}{2}(\boldsymbol{P}-2\boldsymbol{q})\right]~,
\end{equation}
which can be integrated over the angle of $\boldsymbol{q}$ analytically to give a function of $P = |\boldsymbol{P}|$ and $q = |\boldsymbol{q}|$. At this point we find the only dependence on the angle of $\boldsymbol{P}$ is within the final Fermi factors
\begin{equation}
    \left(1-f\left[\frac{1}{2}(\boldsymbol{P}+2q \hat{\boldsymbol{q}}'\right]\right)\left(1-f\left[\frac{1}{2}(\boldsymbol{P}-2q \hat{\boldsymbol{q}}'\right]\right)~,
\end{equation}
where $\hat{\boldsymbol{q}}'$ is a unit vector in the direction of $\boldsymbol{q}'$. As for the initial Fermi factors, this can be integrated over the angle of $\boldsymbol{P}$ analytically to give a function of $P$ and $q$. The final angular integral is trivial since the only angular dependence is now contained within the differential cross section, leaving 
\begin{equation}
    \int d\Omega_{\text{CM}} \left(\frac{d\sigma}{d\Omega}\right)_\text{CM} = \sigma(E_K^\text{CM})~,
\end{equation}
where $E_K^\text{CM} = q^2/m_N$ is the total kinetic energy in the CM-frame. Our phase space integral thus reduces to two integrals, one over $q$ and one over $P$. We change to dimensionless variables through the substitutions
\begin{equation}
    x = \frac{E_K^\text{CM}}{T} = \frac{q^2}{m_N T}\, , \quad y = \frac{P^2}{m_N T}~.
\end{equation}
We then extract the $y$ integral, whose integrand arose from our angular integrals over Fermi factors, into a function $\bar{\Sigma}(x,\eta_1,\eta_2)$, and normalize it to the total number density of nucleons using
\begin{equation}
    n_i = 2\left(\frac{m_N T}{2\pi}\right)^{3/2}\left[-\text{Li}_{3/2}(-e^{-\eta_i})\right]~,
\end{equation}
leaving us with the expression 
\begin{equation} \label{eqn: brem production final}
    \Gamma_{\text{prod/abs}}^{\text{br}}(\omega)=\frac{8\mathcal{S}g_N^2 n_1 n_2 m_\phi^4}{\pi m_N^2}\sqrt{\frac{\pi T}{m_N}}\frac{1}{\omega^5}\int_{x_{\text{min}}}^{\infty} dx \, x \, \sigma(x T) \, \Bar{\Sigma}(x,\eta_1,\eta_2)~,
\end{equation}
where $\bar{\Sigma}$ is given by
\begin{equation}
    \bar{\Sigma}(x,\eta_1,\eta_2) = \frac{1}{4\sqrt{\pi}\text{Li}_{3/2}(-e^{-\eta_1})\text{Li}_{3/2}(-e^{-\eta_2})}\int_0^\infty g(x,y,\eta_1,\eta_2)\; dy~,
\end{equation}
with
\begin{equation}
    g(x,y,\eta_1,\eta_2) = \frac{1}{x\sqrt{y}(\cosh(x+y-\eta_1-\eta_2)-1)}\ln\left[\frac{\cosh\left(\frac{(\sqrt{x}-\sqrt{y})^2-\eta_1-\eta_2}{2}\right)+\cosh\left(\frac{\eta_1-\eta_2}{2}\right)}{\cosh\left(\frac{(\sqrt{x}+\sqrt{y})^2-\eta_1-\eta_2}{2}\right)+\cosh\left(\frac{\eta_1-\eta_2}{2}\right)}\right]^2 \, . 
\end{equation}
The degeneracy parameters $\mu/T$ of the relevant particles are given by $\eta_1$ and $\eta_2$. The normalization of $\bar{\Sigma}$ is such that in the non-degenerate limit, $\eta_i\rightarrow -\infty$, $\bar{\Sigma}(x, \eta_1,\eta_2)\rightarrow e^{-x}$. This function also approaches zero in the completely degenerate limit, up to some value of $x$ corresponding to the Fermi surface, representing the fact that interactions below this energy are completely Pauli-blocked.

\begin{figure}[h]
    \centering
    \includegraphics[width=0.7\linewidth]{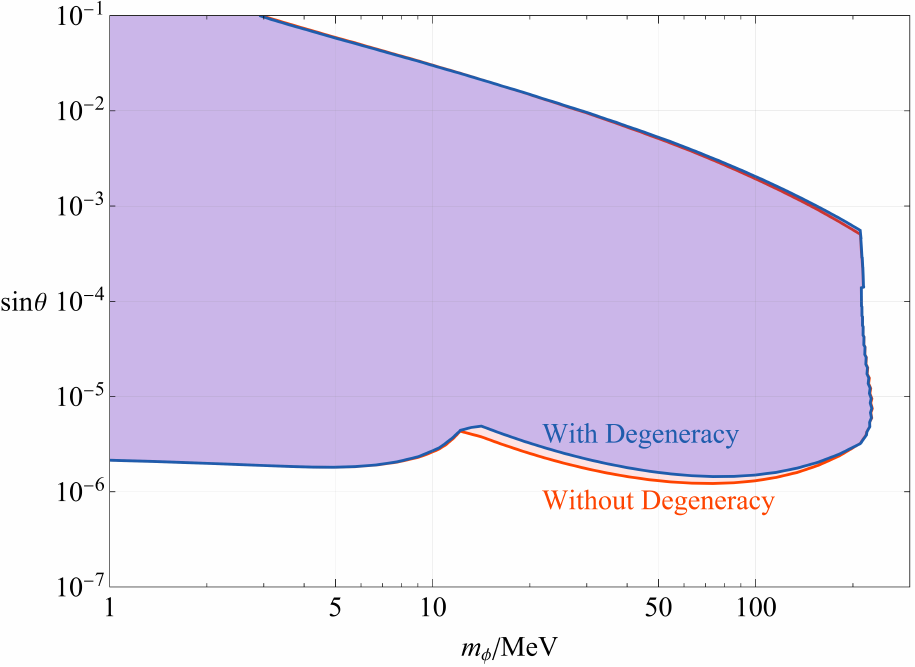}
    \caption{SN~1987A constrains on a Higgs-mixed scalar, obtained with and without the inclusion of degeneracy effects in the calculation of the bremsstrahlung rate. When such corrections are included Pauli-blocking leads to a small decrease in the production rate, which weakens the bounds but only slightly.}
    \label{fig:effects of degeneracy on brem}
\end{figure}

The effect of degeneracy on our final bounds on a Higgs-mixed scalar is shown in Figure~\ref{fig:effects of degeneracy on brem}. Despite the neutrons being partially degenerate in the core, the change in the bounds is minor.

\section{Nucleon recoil effects on bremsstrahlung}\label{appendix: no nucleon recoil}
In Section~\ref{sec: continuum production} we argued that the interplay of Pauli-blocking and the ``no nucleon recoil'' approximation led to the production rate being overestimated and the absorption rate being underestimates (and hence, we made use of detailed balance to obtain conservative limits). Here we repeat the analysis of Appendix~\ref{appendix: degeneracy bremsstrahlung} but retaining the scalar's 4-momentum $K^\mu=(\omega,\boldsymbol{k})$ in the delta function to justify these claims. We begin by making the same change of variables as in Eq.~\eqref{eqn: brem change of vars}, however now conservation of momentum imposes $\boldsymbol{P}' = \boldsymbol{P}-\boldsymbol{k}$, and energy conservation imposes $q' = \tilde{q}$ where 
\begin{equation}
    \tilde{q}^2=q^2-\frac{k^2}{4}+\frac{\boldsymbol{P}\cdot\boldsymbol{k}}{4}-m_N \omega.
\end{equation}
We next identify the angle of $\boldsymbol{q}$ with the CM-frame angle $d\Omega_q=d\Omega_\text{CM}$\footnote{This differs from the choice in Appendix~\ref{appendix: degeneracy bremsstrahlung}, since now there is an ambiguity in the final state as the CM-frame of the 4-nucleon system no longer coincides with the CM-frame of the full system including the scalar.}, allowing us to extract the integral over the angle of $\boldsymbol{q}'$ 
\begin{align}
    I_1(P,q,\boldsymbol{P}\cdot\boldsymbol{k}) &= \int d\Omega_{q'} \left(1-f\left[\frac{\boldsymbol{P}-\boldsymbol{k}+2 \tilde{q}\boldsymbol{\hat{q}'}}{2}\right]\right)\left(1-f\left[\frac{\boldsymbol{P}-\boldsymbol{k}-2 \tilde{q}\boldsymbol{\hat{q}'}}{2}\right]\right)\\
    &=\frac{2\pi}{b (1 - e^{\eta_1 + \eta_2 - 2a})} \log \left( \frac{(e^{a+b} + e^{ \eta_1})(e^{a+b} + e^{ \eta_2})}{(e^{a} + e^{b + \eta_1})(e^{a} + e^{b + \eta_2})} \right)
\end{align}
where 
\begin{equation}
    a = \frac{1}{8m_N T}(P^2+4q^2-4m_N \omega)
\end{equation}
and
\begin{equation}
   b = \frac{1}{8m_N T}(4\tilde{q}|\boldsymbol{P}-\boldsymbol{k}|).
\end{equation}
To proceed, we find it useful to perform an average over the angle of $\boldsymbol{k}$ by integrating $\frac{1}{4\pi}\int d\Omega_k$, which will not change our production rate since the medium is isotropic. This allows us to extract the following dimensionless function
\begin{equation}
    \tilde{J}(P,q) = \int d\Omega_{k}\frac{2 \tilde{q}(q,\boldsymbol{P}\cdot\boldsymbol{k_s})}{m_N}I_1(P,q,\boldsymbol{P}\cdot\boldsymbol{k_s}),
\end{equation}
which must be evaluated numerically. The integral over the angle of $\boldsymbol{P}$ can now be performed analytically to give
\begin{align}
    I_2(P,q) &= \int d\Omega_P f\left[\frac{\boldsymbol{P}+2\boldsymbol{q}}{2}\right]f\left[\frac{\boldsymbol{P}-2\boldsymbol{q}}{2}\right],\\
    &=\frac{2\pi}{(1 - e^{x + y - \eta_1 - \eta_2}) \sqrt{xy}} \log \left( \frac{\cosh \left( \frac{(\sqrt{x} - \sqrt{y})^2 - \eta_1 - \eta_2}{2} \right) + \cosh \left( \frac{\eta_1 - \eta_2}{2} \right)}{\cosh \left( \frac{(\sqrt{x} + \sqrt{y})^2 - \eta_1 - \eta_2}{2} \right) + \cosh \left( \frac{\eta_1 - \eta_2}{2} \right)} \right).
\end{align}
When all of this is assembled, we reach an expression for the production rate identical to Eq.~\eqref{eqn: brem production final}, except with the replacement $g(x,y,\eta_1,\eta_2)\rightarrow G(x,y,w,\delta,\varepsilon,\eta_1,\eta_2)$ where
\begin{equation}
    G(x,y,w,\delta,\varepsilon,\eta_1,\eta_2) = \frac{1}{16\pi^3\sqrt{\delta}}\sqrt{\frac{y}{x}}I_2(x,y,\eta_1,\eta_2)\tilde{J}(x,y,w,\delta,\varepsilon,\eta_1,\eta_2),
\end{equation}
and with the understanding that the integral should be taken over the region satisfying the inequality
\begin{equation}
    x+\sqrt{yw}\sqrt{w\delta-\varepsilon}>w\left(1+\frac{w\delta-\epsilon}{4}\right).
\end{equation}
for the case of production. We now explicitly show the dependence on the dimensionless parameters
\begin{equation}
    w = \frac{\omega}{T},\quad
    \delta = \frac{T}{m_N},\quad
    \varepsilon = \frac{m_\phi^2}{m_N\omega}.
\end{equation}
It can be shown that in the soft limit $w\rightarrow 0$, this reduces back to the result of the no nucleon recoil calculation $g(x,y,\eta_1,\eta_2)$, and in the limit $\delta,\varepsilon\rightarrow 0$ the integration region reduces to $x>w$. 

\begin{figure}
    \centering
    \includegraphics[width=0.7\linewidth]{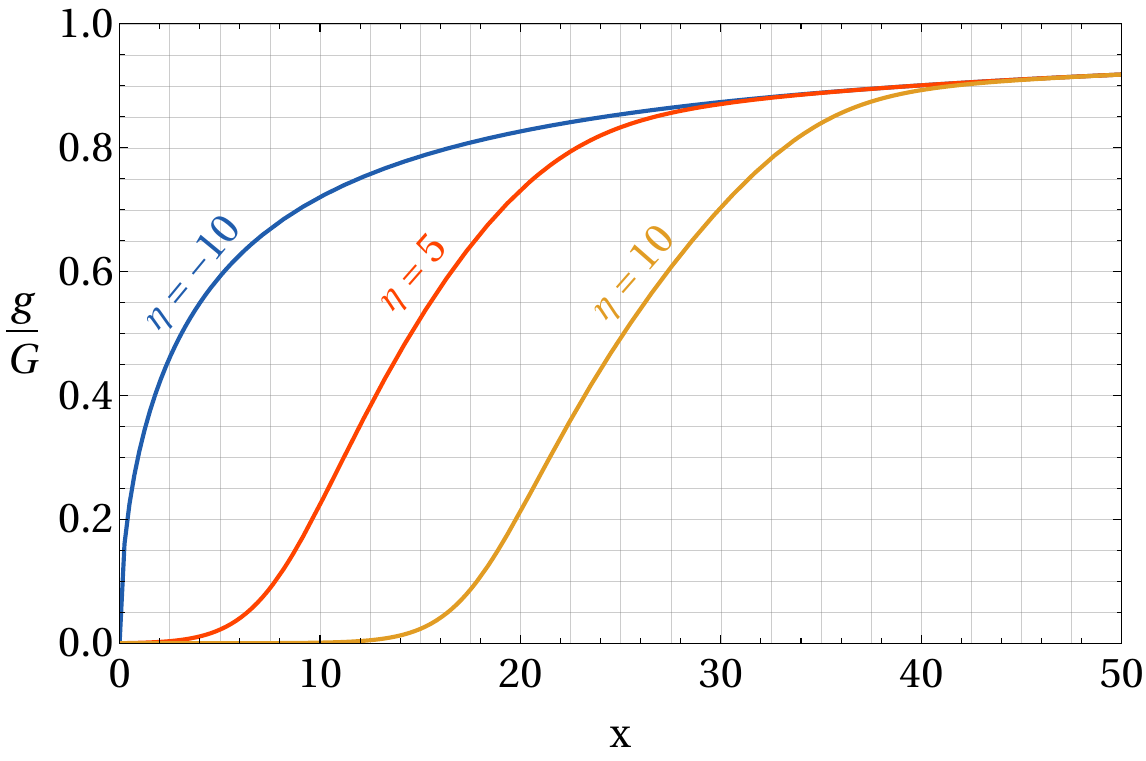}
    \caption{The ratio of the function $g$ entering in the no nucleon recoil approximation for inverse bremsstrahlung absorption with its corrected value $G$, plotted against the energy in the CM-frame. We take $T = 30~{\rm MeV}$, $m_\phi = 100~{\rm MeV}$, $\omega = 300~{\rm MeV}$ and $y=2$ corresponding to typical values inside the supernova core. We take the two chemical potentials to be the same in this case. $\eta = -10$ corresponds to non-degenerate nucleons, $\eta = 5$ is a typical value for the very core of the supernova, and $\eta = 10$ corresponds to degenerate nucleons.}
    \label{fig:g vs G absorption}
\end{figure}

Numerical evaluation shows that $G(x,y,w,\delta,\varepsilon,\eta_1,\eta_2)<g(x,y,\eta_1,\eta_2)$ for physically relevant parameter values, and since the increase in integration region is relatively small, we expect that accounting for nucleon recoils does decrease the production rate, in line with our discussion of Section~\ref{sec: continuum production}. The comparison for absorption is more straightforward since the integration region is remains $x>0,y>0$, so the replacement of $g(x,y,\eta_1,\eta_2)$ with $G(x,y,-w,\delta,-\varepsilon,\eta_1,\eta_2)$ is the only effect of including nucleon recoil\footnote{The negative signs here arise from $K^\mu$ appearing with the opposite sign in the 4-momentum-conserving delta function.}. We therefore plot the ratio $g(x,y,\eta_1,\eta_2)/G(x,y,-w,\delta,-\varepsilon,\eta_1,\eta_2)$ in Figure~\ref{fig:g vs G absorption} which shows that  $G(x,y,-w,\delta,-\varepsilon,\eta_1,\eta_2) > g(x,y,\eta_1,\eta_2)$ and therefore neglecting nucleon recoil underestimates the absorption rate as predicted in Section~\ref{sec: continuum production}. Our approach of calculating first the absorption rate and then using it to find the production rate via detailed balance is therefore conservative.

Finally, we have also checked the rough approximation to the production and absorption rates obtained by taking the simplified integration region $x>w$ and assuming that $\sigma(xT)$ is constant. In this case, we find that, as expected, accounting for nucleon recoils leads to a much better agreement with the predictions of detailed balance compared with the no nucleon recoil approximation, which as discussed in Section~\ref{sec: continuum production} led to a discrepancy of multiple orders of magnitude.

\section{Details of supernova profiles}\label{appendix: sn profiles}

\begin{figure}
    \centering
    \subfloat
    {{\includegraphics[width=0.475\textwidth]{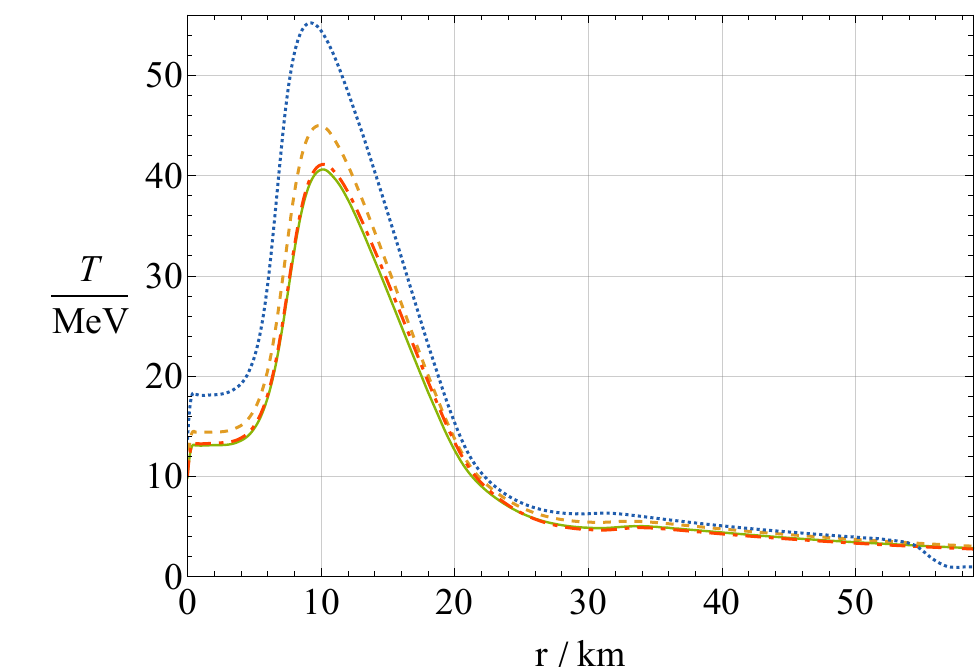} }} \quad
    \subfloat
    {{\includegraphics[width=0.475\textwidth]{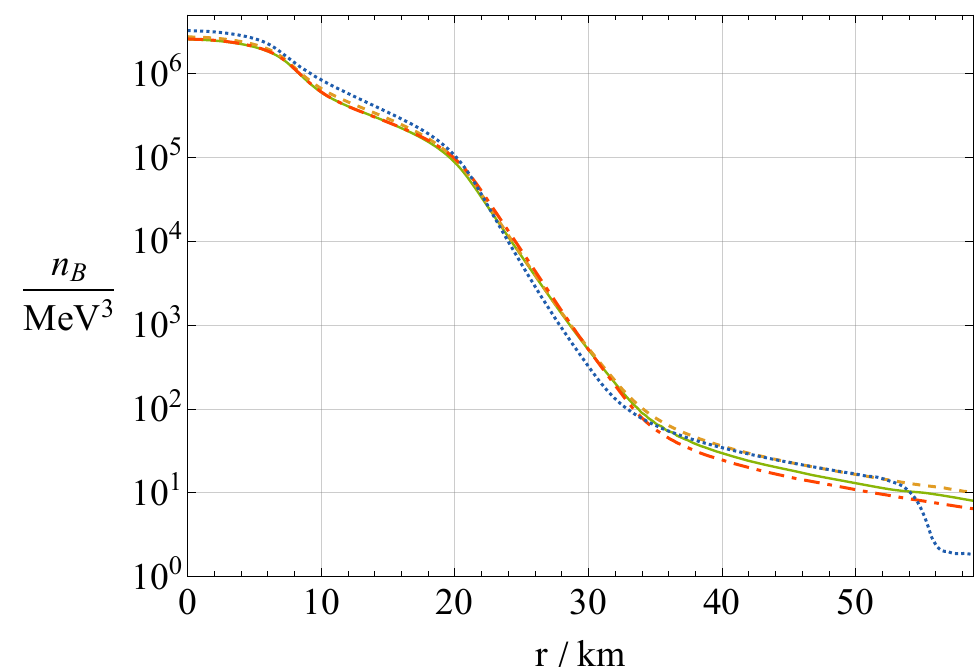} }}\\
    \subfloat
    {{\includegraphics[width=0.475\textwidth]{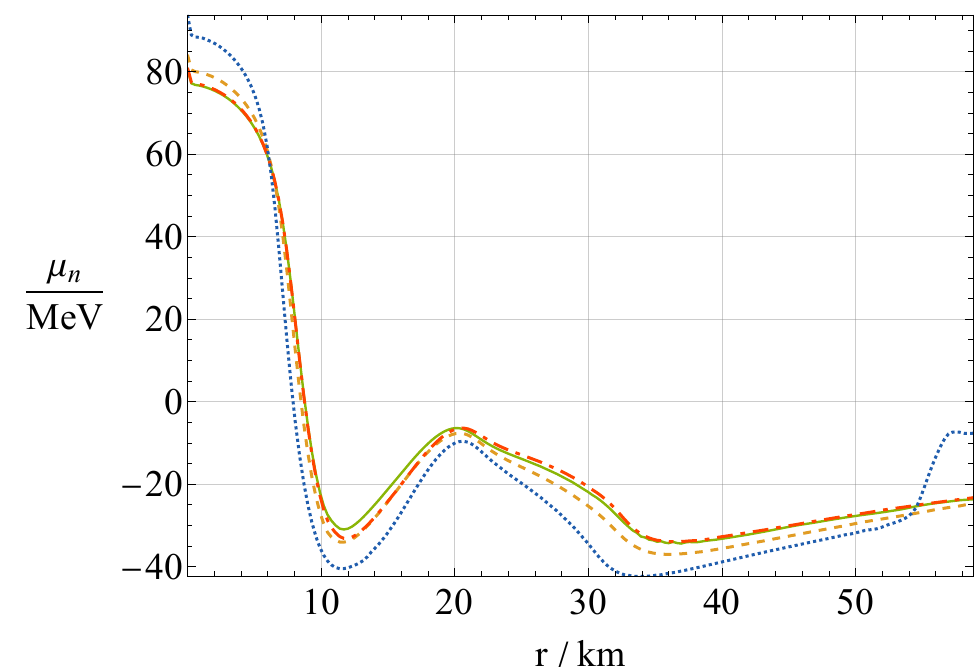} }} \quad
    \subfloat
    {{\includegraphics[width=0.475\textwidth]{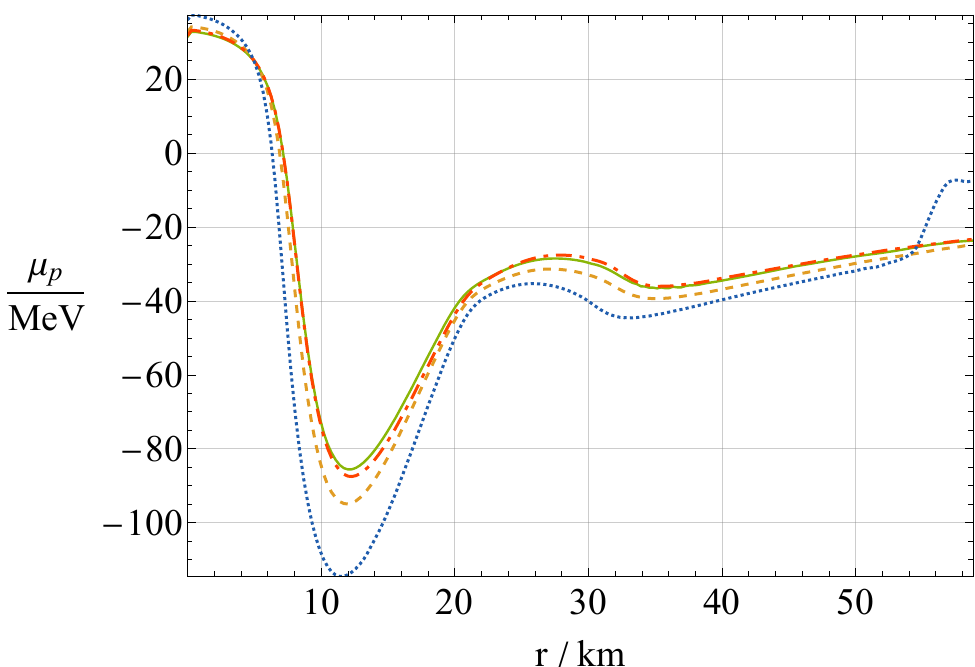} }}\\
    \subfloat
    {{\includegraphics[width=0.475\textwidth]{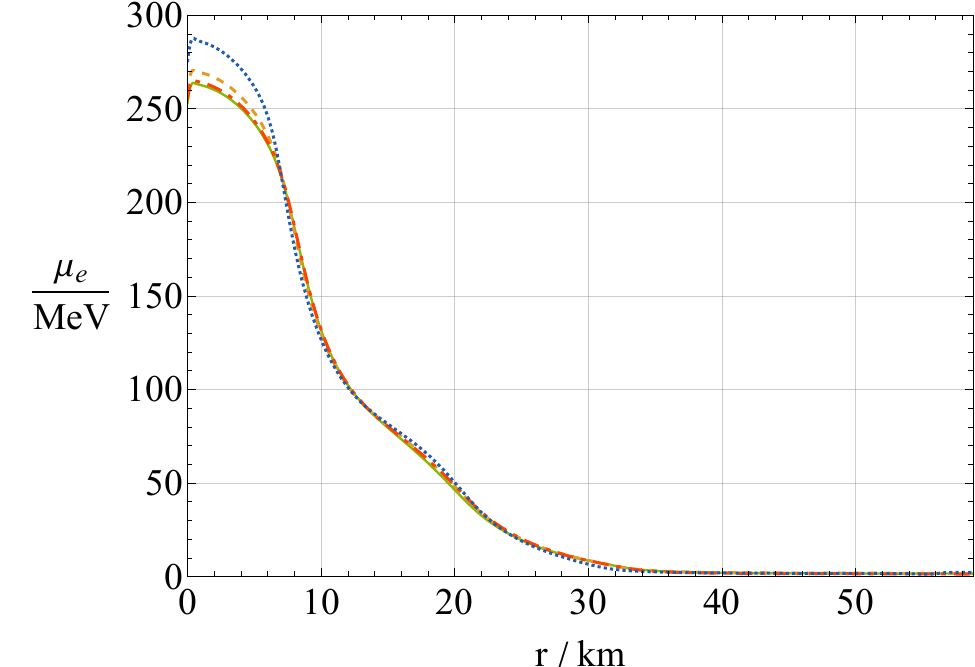} }} \quad
    \subfloat
    {{\includegraphics[width=0.475\textwidth]{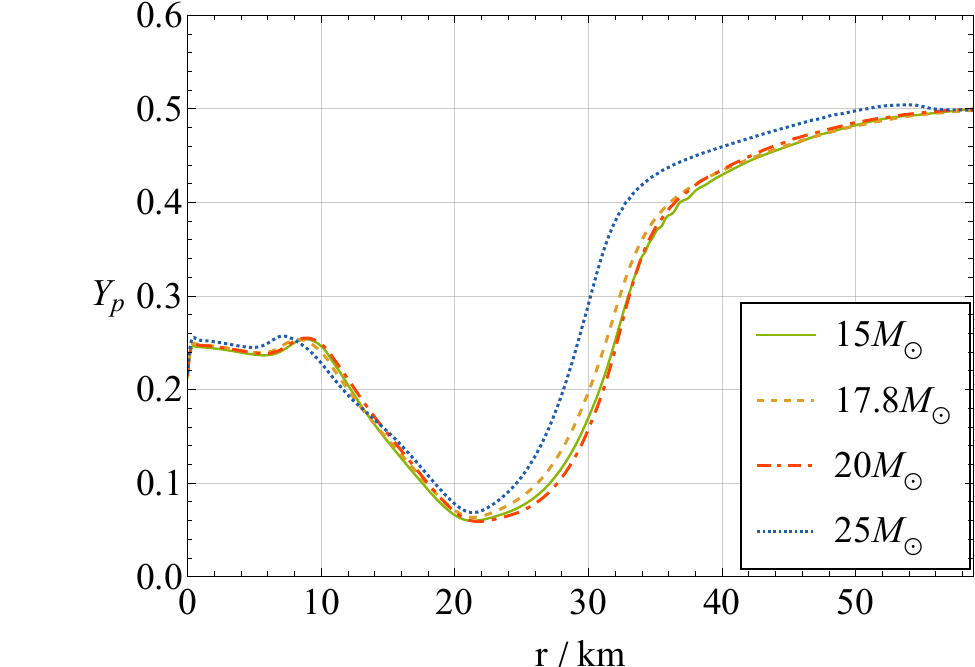} }}
    \caption{The temperature $T$, baryon number density $n_B$, neutron chemical potential $\mu_n$, proton chemical potential $\mu_p$, electron chemical potential $\mu_e$, and proton fraction $Y_p$ as a function of radius from the centre of the supernova $r$ for the four different progenitors that we use. The neutrinosphere radius is $R_{\nu} \simeq 30$~km for all the progenitors.} 
    \label{Fig: SN profiles} 
\end{figure}

The supernova profiles that we consider are plotted in Figure~\ref{Fig: SN profiles}. As can be seen the different progenitors all result in profiles with broadly similar features. The $25~M_\odot$ model has slightly higher temperatures and electron chemical potential in its core, which is consistent with it leading to slightly stronger constraints in Figure~\ref{fig:Higgs-portal results}.

The effects of degeneracy such as Pauli-blocking are relevant when the nucleon number density is comparable to or larger than $2n_Q$, where the factor of 2 accounts for the number of nucleon spin states and $n_Q$ is the quantum concentration defined as
\begin{equation}
    n_Q = \left(\frac{m_N T}{2\pi}\right)^{3/2}~.
\end{equation}
From Figure~\ref{Fig: SN profiles} we see that near the centre of the protoneutron star (out to around $5~{\rm km}$) the temperature is roughly $15~{\rm MeV}$ for all profiles, which corresponds to a quantum concentration of $n_Q \sim 10^5~{\rm MeV}^3$. Since the neutron number density is over an order of magnitude larger than $n_Q$, we expect to be in the regime where quantum effects such as Pauli-blocking are potentially important. An alternative way to reach the same conclusion is by calculating the neutron chemical potential, which leads to a neutron degeneracy parameter $\eta_n = \mu_n/T \lesssim 6 $. The neutron number density drops below the quantum concentration at around $10~{\rm km}$, such that outside of this radius they are non-degenerate. We can also see that electrons are degenerate all the way out to around $30~{\rm km}$, i.e. roughly to the neutrinosphere radius $R_{\nu}$. Between $R_{\nu}$ and the shock front, electrons cease being degenerate due to deleptonization processes during the accretion phase.

\newpage

\bibliographystyle{JHEP}
\bibliography{snescalars}

\end{document}